\documentclass[a4paper,fleqn,usenatbib]{mnras}

\usepackage{mathptmx}

\usepackage[T1]{fontenc}
\usepackage{ae,aecompl}

\usepackage{graphicx}	
\usepackage{amsmath}	
\usepackage{amssymb}	

\usepackage{graphicx,lscape,amssymb,amsmath,url,placeins,mathrsfs,float}

\def\aap{A\hbox{\rm \&}A} 
  \def\aj{AJ} \def\ao{Appl.\ Opt.}
  \def\apj{ApJ} \def\apjl{ApJ}
 \def\apjs{ApJS} 
\def\araa{ARA\hbox{\rm \&}A}

\def\mnras{MNRAS} \def\nat{Nat} \def\pasj{PASJ}

\title[Galaxy far-IR LF out to $\bmath{z \simeq 5}$]{The evolving far-IR galaxy luminosity function and dust-obscured star-formation rate density out to $\bmath{z \simeq 5}$}

\author[M. Koprowski et al.]{
M.\,P. Koprowski,$^{1,2}$\thanks{E-mail: m.koprowski@herts.ac.uk}
J.\,S. Dunlop,$^{2}$
M.\,J. Micha{\l}owski,$^{2,3}$
K.\,E.\,K. Coppin,$^{1}$
\newauthor
\,\,J.\,E. Geach,$^{1}$
\,\,R.\,J. McLure,$^{2}$
\,\,D. Scott,$^{4}$ and P.\,P. van der Werf\,$^{5}$
\\
$^{1}$Centre for Astrophysics Research, Science \& Technology Research Institute, University of Hertfordshire, Hatfield AL10 9AB, UK\\
$^{2}$SUPA\thanks{Scottish Universities Physics Alliance}, Institute for Astronomy, University of Edinburgh, Royal Observatory, Edinburgh, EH9 3HJ, UK\\
$^{3}$Astronomical Observatory Institute, Faculty of Physics, Adam Mickiewicz University, ul.~S{\l}oneczna 36, 60-286 Pozna{\'n}, Poland\\
$^{4}$Department of Physics and Astronomy, 6224 Agricultural Road, University of British Columbia, Vancouver V6T 1Z1, Canada\\
$^{5}$Leiden Observatory, Leiden University, PO Box 9513, NL-2300 RA Leiden, the Netherlands
}

\date{Accepted XXX. Received YYY; in original form ZZZ}

\pubyear{2017}

\begin{document}
\label{firstpage}
\pagerange{\pageref{firstpage}--\pageref{lastpage}}
\maketitle

\begin{abstract}
We present a new measurement of the evolving galaxy far-IR luminosity function (LF) extending out to redshifts $z \simeq 5$, with resulting implications for the level of dust-obscured star-formation density in the young Universe. To achieve this we have exploited recent advances in sub-mm/mm imaging with SCUBA-2 on the James Clerk Maxwell Telescope (JCMT) and the Atacama Large Millimeter/Submillimeter Array (ALMA), which together provide unconfused imaging with sufficient dynamic range to provide meaningful coverage of the luminosity-redshift plane out to $z>4$. Our results support previous indications that the faint-end slope of the far-IR LF is sufficiently flat that comoving luminosity-density is dominated by bright objects ($\simeq L^*$). However, we find that the number-density/luminosity of such sources at high redshifts has been severely over-estimated by studies that have attempted to push the highly-confused {\it Herschel} SPIRE surveys beyond $z \simeq 2$. Consequently we confirm recent reports that cosmic star-formation density is dominated by UV-visible star formation at $z > 4$. Using both direct ($1/V_{\rm max}$) and maximum likelihood determinations of the LF, we find that its high-redshift evolution is well characterized by continued positive luminosity evolution coupled with negative density evolution (with increasing redshift). This explains why bright sub-mm sources continue to be found at $z>5$, even though their integrated contribution to cosmic star-formation density at such early times is very small. The evolution of the far-IR galaxy LF thus appears similar in form to that already established for active galactic nuclei, possibly reflecting a similar dependence on the growth of galaxy mass.
\end{abstract}

\begin{keywords}
dust, extinction -- galaxies: evolution, high-redshift, luminosity function, star formation -- cosmology: observations
\end{keywords}

\section{Introduction}

\label{sec:intro}

A key challenge in modern astrophysical cosmology is to complete our knowledge of cosmic star-formation history, taking proper account of dust-obscured activity (e.g. \citealt{Madau_2014}). Achieving this requires a reliable measurement of the form and evolution of both the rest-frame UV {\it and} rest-frame far-IR galaxy luminosity functions (LFs)  out to the highest redshifts. This is because a fair census of both unobscured and dust-obscured comoving star-formation rate density ($\rho_{\rm SFR}$) requires the luminosity-weighted integration of the relevant LFs over sufficient dynamic range to properly account for the contributions of the brightest sources, while at the same time quantifying the impact of the adopted lower luminosity limit for the integration (i.e. reliably establishing the faint-end slope of the LF).

In recent years this goal has been largely achieved at rest-frame UV wavelengths over nearly all of cosmic history, through observations with the {\it Hubble Space Telescope} ({\it HST}\/) and wider-area ground-based imaging from Subaru, CFHT, VLT and VISTA \citep{Cucciati_2012, McLure_2013b, Bowler_2014, Bowler_2015, Bouwens_2015, Bouwens_2016, Finkelstein_2015, Parsa_2016}. Indeed, meaningful disagreements over the comoving UV luminosity density ($\rho_{\rm UV}$) produced by the evolving galaxy population are now largely confined to extreme redshifts $z > 8$ \citep{Ellis_2013, Oesch_2014, McLeod_2015}, and even here much of the claimed disagreement can be removed by the adoption of consistent limits to the LF integration \citep{McLeod_2016, Ishigaki_2017}. The reason that the adopted faint integration limit becomes an issue for calculating $\rho_{\rm UV}$ at extreme redshifts (and the resulting contribution of the emerging early galaxy population to reionization; \citealt{Robertson_2013, Robertson_2015}) is that the faint-end slope ($\alpha$) of the galaxy UV LF steepens with increasing redshift, from relatively modest values at intermediate redshifts (e.g. $\alpha \simeq -1.3$ at $z \simeq 2$; \citealt{Parsa_2016, Mehta_2017}) to $\alpha \simeq -2$ at $z > 7$ \citep{McLure_2013b, Bouwens_2015, McLeod_2015, McLeod_2016}. As a result, despite the discovery of significant numbers of UV-bright galaxies at $z > 6$ \citep{Bowler_2014, Bowler_2015}, $\rho_{\rm UV}$ within the first $\simeq$\,Gyr of cosmic time is dominated by the contributions of the numerous faintest galaxies, and so the derived value depends more critically on how far down in luminosity the LF extends than is the case at later times (see \citealt{Parsa_2016}).

Despite the long-established importance of the far-IR background \citep{Dole_2006}, similar progress in our knowledge of the far-IR LF has been hampered by the heightened observational challenges at mid/far-IR and sub-mm/mm wavelengths (i.e. high background and poor angular resolution) and a resulting lack of survey dynamic range (coupled with uncertainties in redshift content). Nevertheless, important progress has been made over the past decade, first with mid-IR observations using NASA's {\it Spitzer Space Telescope}, and more recently through far-IR imaging with ESA's {\it Herschel Space Observatory}. Together, these facilities have enabled the far-IR LF and its evolution to be successfully traced out to $z \simeq 2$. 

First, {\it Spitzer} MIPS 24\,${\rm \mu m}$ imaging was used to study the mid-IR LF out to $z \simeq 2$ \citep{LeFloch_2005, Caputi_2007, Rodighiero_2010}, albeit extrapolation from 24\,${\rm \mu m}$ to far-IR luminosity becomes increasingly dangerous with increasing wavelength (although see \citealt{Elbaz_2010}). Attempts were also made to exploit the 70\,${\rm \mu m}$ imaging provided by {\it Spitzer} \citep{Magnelli_2009, Patel_2013}, but sensitivity/resolution limitations largely restricted the usefulness of this work to $z < 1$ (although \citealt{Magnelli_2011} pushed out to $z\simeq2$ via stacking). 

Over the last five years, {\it Herschel} PACS and SPIRE surveys have enabled this work to be developed through object selection at more appropriate far-IR wavelengths (i.e. closer to the rest-frame peak of the far-IR emission). \citet{Magnelli_2013} used deep {\it Herschel} PACS  imaging from the PEP and GOODS surveys to determine the bright-end of the far-IR LF out to $z \simeq 2$, while \citet{Gruppioni_2013} used the PEP PACS imaging and SPIRE HerMES imaging (at 250, 350 and 500\,${\rm \mu m}$) to try to extend this work out to $z \simeq 4$ (see also \citealt{Burgarella_2013}). The inclusion of the 250, 350 and 500\,${\rm \mu m}$ data allowed \citet{Gruppioni_2013}, in principle, to determine far-IR luminosities without recourse to large extrapolations, but object selection was still undertaken at the shorter (PACS) wavelengths. As a result, this study is mostly sensitive to the lower-redshift/warmer sources, and hence, at $z > 2$, allows only the detection of the most extreme sources. One consequence of the resulting lack of dynamic range is that the faint-end slope of the far-IR LF could not be measured at high redshift, and so \citet{Gruppioni_2013} simply adopted the $z = 0$ value at all higher redshifts. Most recently, \citet{Rowan_2016} attempted to expand on the \citet{Gruppioni_2013} study, and to extend it out to even higher redshifts ($z \simeq 6$) by including object selection at 500\,${\rm \mu m}$ (the longest-wavelength {\it Herschel} SPIRE imaging band). This study yielded surprisingly high estimates of far-IR luminosity density at high redshifts; however, utilising the longest-wavelength {\it Herschel} data in this way is fraught with danger due to the large beamsize ($\simeq 36$\,arcsec FWHM) and consequent issues regarding blending, source mis-identification, potential AGN contamination and gravitational lensing (e.g. star-formation rates, SFR, as high as 20,000\,${\rm M_{\odot} yr^{-1}}$ are reported in this work).

Thus, despite this impressive progress, it is clear that attempting to reliably measure the far-IR galaxy LF based on object selection in {\it Spitzer} or {\it Herschel} surveys becomes increasingly problematic beyond $z \simeq 2$. It is now possible to overcome these difficulties at high redshift by moving to ground-based sub-mm/mm object selection, using a combination of wide-area imaging surveys as produced by SCUBA-2 \citep{Holland_2013} on the JCMT (e.g. \citealt{Geach_2013, Geach_2017, Roseboom_2013, Chen_2016, Michalowski_2016, Cowie_2017}), and higher-resolution smaller-area mapping as can now be achieved with ALMA (e.g. \citealt{Hatsukade_2015, Hatsukade_2016, Umehata_2015, Walter_2016, Dunlop_2017}). While attempts have previously been made to explore the basic high-redshift evolution of the far-IR population based on ground-based sub-mm data (e.g. \citealt*{Wall_2008}), until now the `wedding-cake' of surveys providing unconfused imaging over a reasonable dynamic range was not of sufficient quality to enable a detailed investigation of the form and evolution of the far-IR LF.

The merits of moving to ground-based sub-mm (i.e. 450 and 850\,${\rm \mu m}$) and mm (i.e. 1.1--1.3\,mm) selection become increasingly obvious as one moves to higher redshifts. First, the smaller beam sizes at these wavelengths offered by large ground-based single-dish telescopes such as the JCMT, or interferometric arrays such as ALMA, substantially reduce/remove the confusion limitations of the longer-wavelength {\it Herschel} SPIRE imaging, minimizing problems of source blending, false counterpart identification, and hence potentially erroneous redshift information. Second, certainly by $z > 3$, object selection at sub-mm/mm wavelengths is much less susceptible to AGN contamination than {\it Spitzer} or {\it Herschel} PACS detections (which sample the spectral energy distribution (SED) shortward of $\lambda_{\rm rest} \simeq 40\,{\rm \mu m}$ at these redshifts), and extrapolation to estimate total far-IR luminosities also becomes less problematic. Third, object selection in the higher-resolution ground-based sub-mm/mm imaging can be used to help deconfuse and deblend the existing {\it Herschel} far-IR imaging, hence enabling its exploitation for SED determination in a less biased way. 

A number of recent studies have already provided indications that the high values of $\rho_{\rm FIR}$ inferred from pushing the {\it Herschel} surveys beyond $z \simeq 2.5$ are incorrect. First, estimates of dust-obscured star-formation activity in known high-redshift galaxies, either derived from analyses of the UV slope (e.g. \citealt{Dunlop_2013}) or from ALMA detections/limits (individual or stacked; e.g. \citealt{Capak_2015, Bouwens_2016b, Koprowski_2016}) suggest that the dust-obscured contribution to $\rho_{\rm SFR}$ at the highest redshifts is relatively small. However, it can reasonably be argued that the most dust-obscured galaxies will not feature in rest-frame UV-selected samples. More significantly, the results from the deep ALMA imaging of the Hubble Ultra Deep Field (HUDF; \citealt{Dunlop_2017}) and the results of stacking in the deepest SCUBA-2 Cosmology Legacy Survey (S2CLS) images \citep{Bourne_2017} both yield much lower values of dust-obscured $\rho_{\rm SFR}$ than derived by \citet{Gruppioni_2013} and \citet{Rowan_2016}, especially at $z > 3$. Nonetheless, it could still be argued that these small-field experiments, while complete to relatively low dust-obscured SFRs ($\simeq 10\,{\rm M_{\odot} yr^{-1}}$), cover insufficient area to reveal the contributions from the most extreme objects. 

In this study we aimed to clarify this situation, and moreover to properly determine the form and high-redshift evolution of the far-IR galaxy LF by analysing the results from the HUDF ALMA 1.3-mm and deep S2CLS 850-${\rm \mu m}$ imaging in tandem with results from the wider-area $850-{\rm \mu m}$ maps recently completed within the S2CLS \citep{Geach_2017, Michalowski_2016}. Together these data provide a sub-mm/mm survey `wedding-cake' with sufficient dynamic range and sufficiently-complete redshift content to enable a meaningful measurement of the form and evolution of the rest-frame 250-${\rm \mu m}$ galaxy LF from the available coverage of the luminosity-redshift plane. Crucially, the bright tier of S2CLS covers over 2 deg$^2$, and yields a sample of $> 1000$ luminous sources (SFR $> 300\,{\rm M_{\odot} yr^{-1}}$), providing excellent sampling of the bright end of the far-IR LF, while the deeper S2CLS imaging and the ALMA imaging enable the first direct measurement of the slope of the faint-end of the far-IR LF at high redshift. Together these data have enabled us to determine the form and evolution of the far-IR LF, and hence the dust-obscured $\rho_{\rm SFR}$ out
to $z \simeq 4.5$.

This paper is structured as follows. The sub-mm/mm imaging utilised in this work, along with the supporting multi-wavelength data, are described in the Section \ref{sec:data}. The multi-wavelength identification process, together with the methods used to establish redshifts for the entire sample are then presented in Section \ref{sec:z}. Next, the procedure for determining the IR LFs, using both the $1/V_{\rm max}$ and the maximum-likelihood methods, is explained in Section \ref{sec:lf}. The resulting calculation of the far-IR and total (UV+far-IR) $\rho_{\rm SFR}$ is presented in Section \ref{sec:sfrd}. We discuss our results in the context of other recent studies in Section \ref{sec:disc}, and conclude by summarising our findings in Section \ref{sec:sum}. 

Throughout the paper we use a \citet{Chabrier_2003} stellar initial mass function (IMF) and assume a flat cosmology with $\Omega_{\rm m} = 0.3$, $\Omega_\Lambda = 0.7$ and H$_0$ = 70 km s$^{-1}$ Mpc$^{-1}$.

\section{Data}
\label{sec:data}

\subsection{JCMT SCUBA-2 imaging}
\label{sec:scuba}

We used the  data collected as a part of the SCUBA-2 Cosmology Legacy Survey (S2CLS). The map-making process and the resulting derived source catalogues are described in \citet{Geach_2017}. The fields utilised here are the UKIDSS-UDS, where the  850-${\rm \mu m}$ imaging covers $\simeq 0.9$\,deg$^2$ with a 1-$\sigma$ noise of 0.9\,mJy (revealing 1085 sources with a signal-to-noise ratio SNR $> 3.5$),  and the COSMOS field, where the 850-${\rm \mu m}$ imaging covers $\simeq 1.3$\,deg$^2$ with the $1\sigma$ noise of 1.6\,mJy (revealing 719 sources with SNR $> 3.5$). These source catalogues are discussed further in \citet{Chen_2016} and \citet{Michalowski_2016}. For simplicity we reduced the effective area of these maps to regions of uniform noise, and for our analysis retained only sources with a simulated completeness $>0.5$. The resulting refined effective survey areas, flux-density limits, SNR, and sample sizes are summarized in Table \ref{tab:submmdata}.

\begin{table}
\begin{normalsize}
\begin{center}
\caption{The refined SCUBA-2 850-${\rm \mu m}$ survey fields and source samples used in this work (see text for details). The columns show the field name, the area, limiting flux density, limiting signal-to-noise ratio and the resulting number of detections. }
\label{tab:submmdata}
\setlength{\tabcolsep}{1.7 mm} 
\begin{tabular}{lcccc}
\hline
            & Area          & $S_{\rm lim}$ & SNR & N     \\
            & /deg$^{-2}$   & /mJy          &     &       \\
\hline
COSMOS deep & $0.08$        & $4.60$        & 4.2 & 34    \\
COSMOS wide & $0.79$        & $6.50$        & 4.3 & 89    \\
UDS         & $0.71$        & $3.75$        & 4.2 & 454   \\
\hline
\end{tabular}
\end{center}
\end{normalsize}
\end{table}

\subsection{ALMA imaging}
\label{sec:alma}

To help inform the measurement of the faint-end slope of the LF, we used the ALMA 1.3-mm imaging of the HUDF undertaken by \citet{Dunlop_2017}. A mosaic of 45 ALMA pointings was created to cover the full $\simeq 4.5$\,arcmin$^2$ area previously imaged with WFC3/IR on {\it HST}. The ALMA map reached a noise level of $\sigma_{1.3}\simeq 35\,{\rm \mu Jy\, beam^{-1}}$, and 16 sources were detected with flux densities $S_{1.3}>120\,{\rm \mu Jy}$. 13 of the 16 sources have spectroscopic redshifts and the remaining three have accurate photometric redshifts derived from the optical--near-IR photometry (for the sources of the spectroscopic redshifts see table 2 in \citealt{Dunlop_2017}).

\subsection{Ancillary data}
\label{sec:auxdata}

For the S2CLS COSMOS field the optical to mid-IR data consist of imaging from the Canada-France-Hawaii Telescope Legacy Survey (CFHTLS; \citealt{Gwyn_2012}), as described by \citet{Bowler_2012}, Subaru \citep{Taniguchi_2007}, the {\it HST} Cosmic Assembly Near-infrared Deep Extragalactic Legacy Survey (CANDELS; \citealt{Grogin_2011}), UltraVISTA Data Release 2 \citep{McCracken_2012, Bowler_2014} and {\it Spitzer} (S-COSMOS; \citealt{Sanders_2007}). At radio wavelengths the VLA-COSMOS Deep \citep{Schinnerer_2010} catalogues were utilised. 

For the S2CLS UDS field the optical data were obtained with Subaru/SuprimeCam \citep{Miyazaki_2002}, as described in \citet{Furusawa_2008}, the near-IR imaging was  provided by the UKIRT Infrared Deep Sky Survey \citep[UKIDSS;][]{Lawrence_2007, Cirasuolo_2010}, the mid-IR data are from the {\it Spitzer} Public Legacy Survey of the UKIDSS Ultra Deep Survey (SpUDS; PI: J. Dunlop), as described in \citet{Caputi_2011}, and the radio (VLA) data are from \citet{Ivison_2005, Ivison_2007} and Arumugam et al. (in preparation). 

For both the UDS and COSMOS fields, far-IR imaging from {\it Herschel} \citep{Pilbratt_2010} was utilised, as provided by the public releases of the HerMES \citep{Oliver_2012} and PEP \citep{Lutz_2011} surveys undertaken with the SPIRE \citep{Griffin_2010} and PACS \citep{Poglitsch_2010} instruments.

For the extraction of the far-IR flux densities and limits, the {\it Herschel} maps at 100, 160, 250, 350 and 500\,${\rm \mu m}$ were utilised, with beam sizes of 7.4, 11.3, 18.2, 24.9, and 36.3\,arcsec, and 5$\sigma$ sensitivities of 7.7, 14.7, 8.0, 6.6, and 9.5\,mJy, respectively. The {\it Herschel} flux densities of (or upper limits for) the SCUBA-2 sources were obtained in the following way. Square image cut-outs of width 120\,arcsec were extracted from each {\it Herschel} map around each SCUBA-2 source, and the PACS (100 and 160\,${\rm \mu m}$) maps were used to simultaneously fit Gaussians with the FWHM of the respective imaging, centred at all radio and 24-${\rm \mu m}$ sources located within these cut-outs, and at the positions of the SCUBA-2 optical identifications (IDs, or just submm positions if no IDs were selected). Then, to deconfuse the SPIRE (250, 350 and 500\,${\rm \mu m}$) maps in a similar way, the positions of the 24\,${\rm \mu m}$ sources detected with PACS (at $> 3\sigma$) were retained, along with the positions of all radio sources, and the SCUBA-2 IDs (or, again, the SCUBA-2 position in the absence of any optical or radio ID).

\section{Redshifts}
\label{sec:z}

\subsection{Multi-wavelength identifications}
\label{sec:ids}

\begin{figure}
\begin{center}
\includegraphics[scale=0.8]{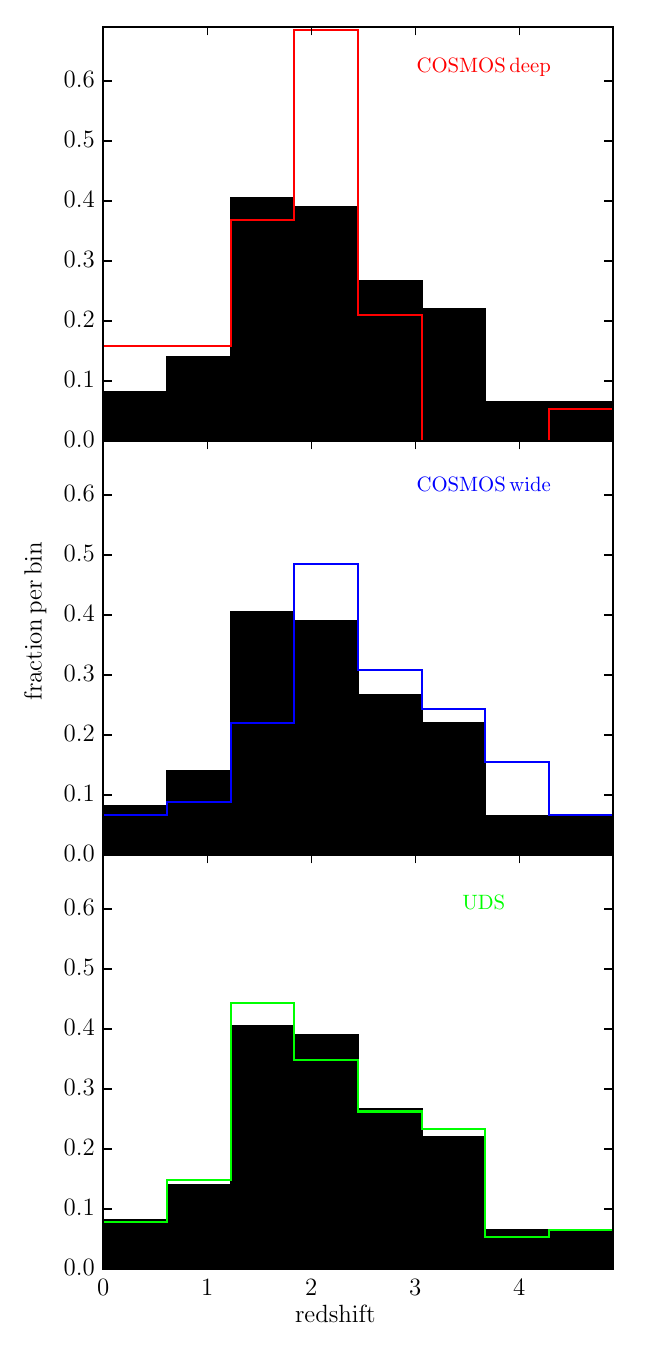}
\end{center}
\caption{The redshift distributions of the refined SCUBA-2 source samples used in this work. The black histogram depicts the distribution for all the sources, and yields a mean redshift of $\bar{z}=2.73\pm 0.06$. From the top, the colour plots show the COSMOS deep, COSMOS wide and the UDS redshift distributions with $\bar{z}=2.30\pm 0.23$, $\bar{z}=3.05\pm 0.17$ and $\bar{z}=2.70\pm 0.07$ respectively.}
\label{fig:z}
\end{figure}

Because of the beam size of the JCMT SCUBA-2 imaging at 850\,${\rm \mu m}$ (FWHM $\simeq 15$ arcsec), we cannot simply adopt the closest optical/near-IR neighbour as the 850-${\rm \mu m}$ source galaxy counterpart. Instead, we used the method outlined in \citet{Downes_1986} (see their Section 5 for a derivation), where we adopt a $2.5\sigma$ search radius around the SCUBA-2 position based on the signal-to-noise ratio (SNR): $r_{\rm s}=2.5\times 0.6\times \rm{FWHM}/\rm{SNR}$. In order to account for systematic astrometry shifts (caused by pointing inaccuracies and/or source blending; e.g. \citealt{Dunlop_2010}) we enforced a minimum search radius of 4.5\,arcsec. Within this radius we calculated the corrected Poisson probability, $p$, that a given counterpart could have been selected by chance.

Three imaging wavebands were used when searching for galaxy counterparts: the VLA 1.4-GHz imaging, the {\it Spitzer} MIPS 24-$\mu$m imaging, and the {\it Spitzer} IRAC 8-$\mu$m imaging. The radio band traces recent star formation via synchrotron radiation from relativistic electrons produced within supernovae (SNe; \citealt{Condon_1992}), whereas the 24-$\mu$m waveband is sensitive to the emission from warm dust. Therefore, since submm-selected galaxies are dusty, highly-star-forming objects, they are expected to be very luminous in these bands. Also, at the redshifts of interest, the 8-${\rm \mu m}$ waveband traces the rest-frame near-IR light coming from the older, mass-dominant stellar populations in galaxies, and thus provides a proxy for stellar mass. Given the growing evidence that sub-mm galaxies are massive, it is expected that 850-${\rm \mu m}$ sources will have significant 8-${\rm \mu m}$ fluxes (e.g. \citealt{Dye_2008, Michalowski_2010, Biggs_2011, Wardlow_2011}). Moreover, the surface density of sources in these three wavebands is low enough for chance positional coincidences to be rare (given a sufficiently small search radius). Once the counterparts were found in each of these bands, they were matched with the optical/near-IR catalogues using a search radius of $r=1.5$ arcsec and the closest object taken to be the galaxy counterpart (for a more detailed description of this process and the relevant identification tables see \citealt{Michalowski_2016}).

\subsection{Redshift distributions}
\label{sec:zdist}

We used the available multi-wavelength data to derive the optical/near-IR photometric redshifts for the SCUBA-2 source galaxy counterparts using a $\chi^2$ minimisation method (\citealt{Cirasuolo_2007, Cirasuolo_2010}) with a code based on the H{\small YPER}Z package \citep{Bolzonella_2000}. To create templates of galaxies, the stellar population synthesis models of \citet{Bruzual_2003} were applied, using the \citet{Chabrier_2003} stellar initial mass function (IMF), with a lower and upper mass cut-off of $0.1$ and $100\,{\rm M_{\odot}}$ respectively. Single and double-burst star-formation histories with a fixed solar metallicity were used. Dust reddening was taken into account using the \citet{Calzetti_2000} law within the range $0 \leq A_V\leq 6$. The HI absorption along the line of sight was applied according to \citet{Madau_1995}. The accuracy of the photometric catalogue of \citet{Cirasuolo_2010} is excellent, with a mean $|z_{\rm phot} - z_{\rm spec}|/(1 + z_{\rm spec}) = 0.008 \pm 0.034$. Also, for every source the `long-wavelength' photometric redshift (e.g. \citealt{Aretxaga_2007, Koprowski_2014, Koprowski_2016b}) was calculated using the SCUBA-2 and {\it Herschel} data by fitting the average sub-mm galaxy SED template of \citet{Michalowski_2010}.

In addition, as explained in detail in \citet{Koprowski_2014, Koprowski_2016b}, we further tested the robustness of our optical identifications by comparing the optical photometric redshifts with the `long-wavelength' photometric redshifts. Sources with optical redshifts that transpired to be significantly lower than their `long-wavelength' ones were considered to be incorrectly identified in the optical (foreground galaxies, possibly lenses) and hence their `long-wavelength' photometric redshifts were adopted (see \citealt{Michalowski_2016} for details).

The final redshift distributions (100\% complete) for the refined fields used in this work (Table \ref{tab:submmdata}) are shown in Figure \ref{fig:z}. Each coloured histogram shows the redshift distribution for the relevant field (as depicted in the legend), with the background black histogram showing the results for all the fields combined. The mean redshifts are $\bar{z}=2.30\pm 0.23$, $\bar{z}=3.05\pm 0.17$ and $\bar{z}=2.70\pm 0.07$ for the COSMOS deep, COSMOS wide and UDS fields respectively. The mean redshift for the whole sample used here is $\bar{z}=2.73\pm 0.06$.

\section{IR Luminosity Function}
\label{sec:lf}

To derive the evolving far-IR LF, $\Phi_{\rm IR}$, we use two independent methods. One is the standard $1/V_{\rm max}$ method \citep{Schmidt_1968}, which allows the calculation of the LFs to be performed directly from the data, without any assumptions regarding the functional shape. To derive the functional form we then fit a set of Schechter functions, where the best-fitting parameter values are found by minimising $\chi^2$. In order to find the continuous form of the redshift evolution of the far-IR LF, we additionally use the maximum-likelihood method presented in \citet{Marshall_1983}, where we again take the LFs to be of a Schechter form.

\subsection{$\pmb{1/V_{\rm max}}$ method}
\label{sec:meth1}

\begin{figure}
\begin{center}
\includegraphics[scale=0.8]{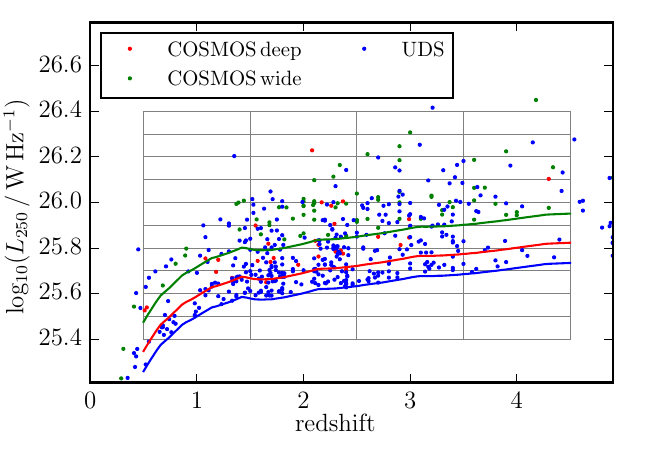}
\end{center}
\caption{The coverage of the luminosity-redshift plane provided by the source samples used in this work. The grey solid lines depict the redshift and luminosity bins used in the LF analysis. The solid colour lines show the corresponding luminosity limits, resulting from the detection limits in each field (see Section \ref{sec:scuba} and Table \ref{tab:submmdata}). These limits are crucial for determining what is the lowest luminosity in each redshift bin at which our sample is complete. Only the luminosity bins for which the minimum luminosity is higher than the luminosity limit are included in the analysis.}
\label{fig:lzp}
\end{figure}

\begin{figure*}
\begin{center}
\includegraphics[scale=0.7]{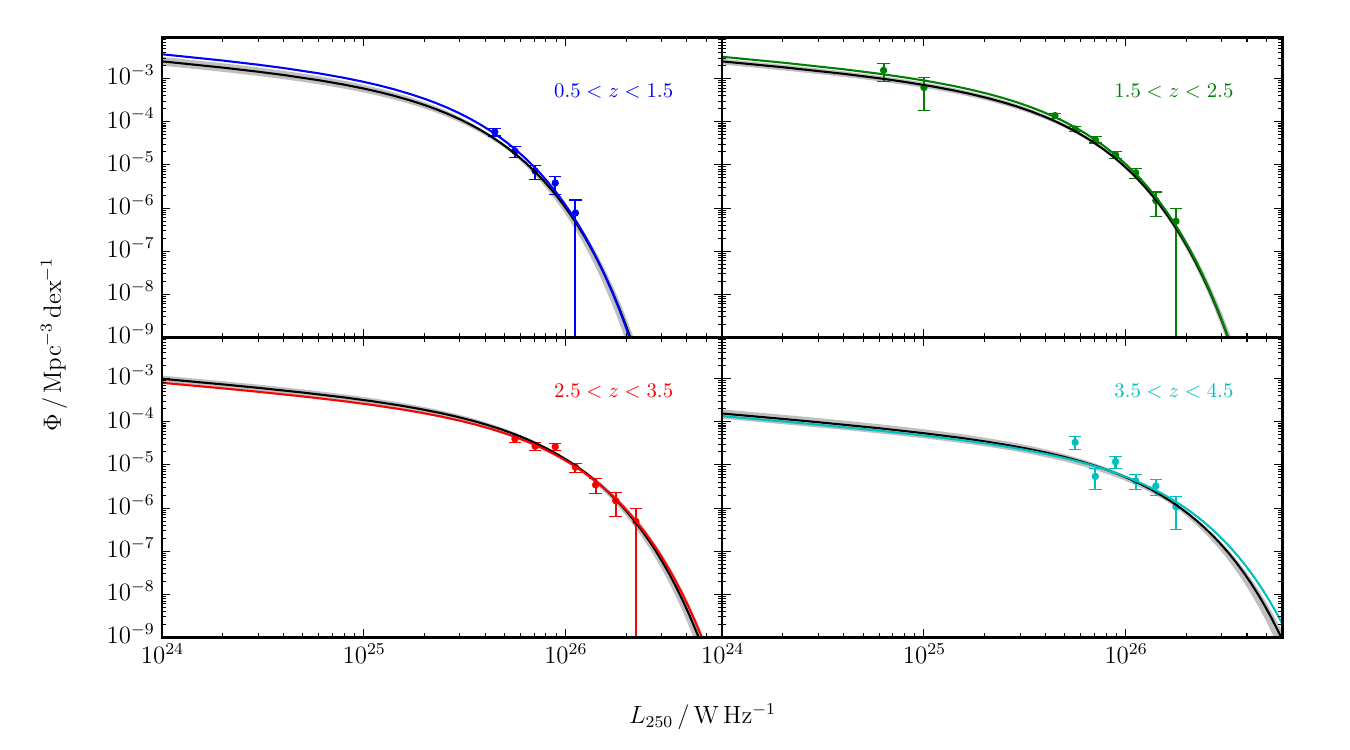}
\end{center}
\caption{The far-IR (rest-frame 250-${\rm \mu m}$) galaxy luminosity functions (LFs) for the four redshift bins studied in this work. The points with error bars show the LF values determined using the $1/V_{\rm max}$ method. The two faintest points in the $1.5<z<2.5$ redshift bin depict the LF values found using the ALMA data. These allowed us to determine the faint-end slope $\alpha=-0.4$, which was then adopted for the other redshift bins. The coloured solid lines show the best-fitting Schechter functions to these data points. The black solid lines (almost perfectly aligned with the coloured solid lines) depict the results of the maximum-likelihood method, with the derived uncertainty indicated by the shaded grey region.}
\label{fig:lf1}
\end{figure*}

\begin{figure*}
\begin{center}
\includegraphics[scale=0.7]{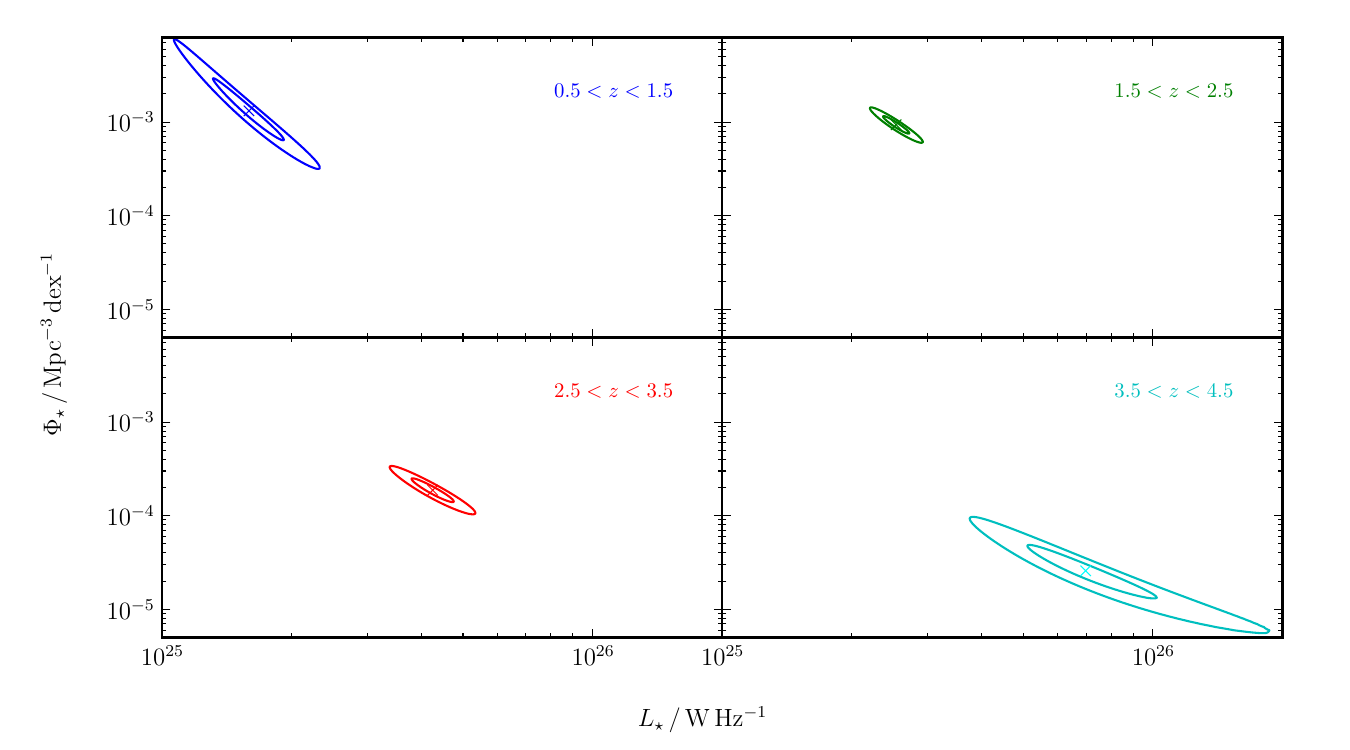}
\end{center}
\caption{The best-fitting parameter values of the Schechter functions found using the $1/V_{\rm max}$ method for the four redshift bins studied in this work. The contours show $\Delta \chi^2 =1$ and $\Delta \chi^2 = 4$ above the best-fitting solution, and hence the corresponding $1\sigma$ and $2\sigma$ uncertainties in the individual parameter values.}
\label{fig:dlf1}
\end{figure*}

\begin{table*}
\begin{footnotesize}
\begin{center}
\caption{Rest-frame 250-${\rm \mu m}$ luminosity functions.}
\label{tab:lf}
\setlength{\tabcolsep}{4 mm} 
\begin{tabular}{ccccc}
\hline
\hline
log($L_{250}/{\rm W\, Hz^{-1}}$) & \multicolumn{4}{c}{log($\Phi/{\rm Mpc^{-3}\, dex^{-1}}$)} \\
{}     & $0.5<z<1.5$     & $1.5<z<2.5$     & $2.5<z<3.5$     & $3.5<z<4.5$     \\
\hline
$24.8$ & ...             & $-2.81\pm 0.16$ & ...             & ...             \\
$25.0$ & ...             & $-3.20\pm 0.23$ & ...             & ...             \\
$25.6$ & $-4.23\pm 0.08$ & $-3.86\pm 0.05$ & ...             & ...             \\
$25.7$ & $-4.68\pm 0.11$ & $-4.17\pm 0.05$ & $-4.40\pm 0.08$ & $-4.47\pm 0.12$ \\
$25.8$ & $-5.14\pm 0.13$ & $-4.42\pm 0.06$ & $-4.56\pm 0.08$ & $-5.27\pm 0.18$ \\
$25.9$ & $-5.42\pm 0.16$ & $-4.76\pm 0.07$ & $-4.58\pm 0.07$ & $-4.93\pm 0.11$ \\
$26.0$ & $-6.11\pm 0.30$ & $-5.18\pm 0.11$ & $-5.05\pm 0.09$ & $-5.37\pm 0.14$ \\
$26.1$ & ...             & $-5.83\pm 0.20$ & $-5.46\pm 0.14$ & $-5.49\pm 0.15$ \\
$26.2$ & ...             & $-6.31\pm 0.30$ & $-5.83\pm 0.20$ & $-5.97\pm 0.23$ \\
$26.3$ & ...             & ...             & $-6.31\pm 0.30$ & ...             \\
\hline
\end{tabular}
\end{center}
\end{footnotesize}
\end{table*}

\begin{figure}
\begin{center}
\includegraphics[scale=0.77]{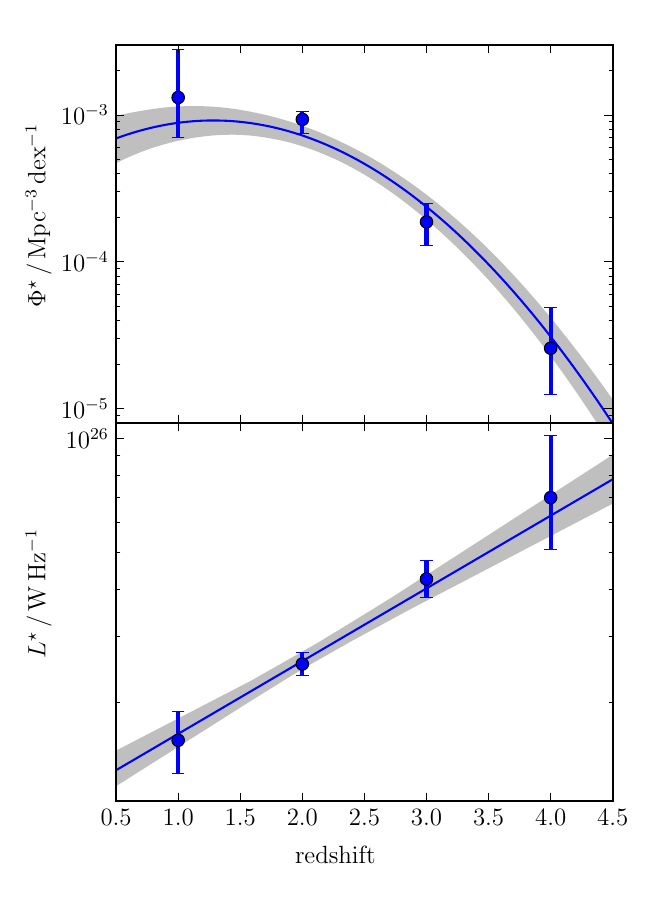}
\end{center}
\caption{Best-fitting Schechter-function parameters. The blue circles depict the best-fitting values determined using the $1/V{\rm max}$ method (Table\,\ref{tab:schparams}), with the $1\sigma$ errors derived by projecting the contours from Fig.\,\ref{fig:dlf1} onto the relevant axis. The blue solid line shows the results from the maximum-likelihood method using the functional form given in Equation \ref{eq:params} with the parameter values given in Table\,\ref{tab:params}. The grey-shaded area depicts the $1\sigma$ errors on the functional form based on the maximum-likelihood ratios, corresponding to $\Delta \chi^2=1$.}
\label{fig:params}
\end{figure}

The LF in a given luminosity and redshift bin is calculated using:

\begin{equation}\label{eq:lf} \Phi(L,z)=\frac{1}{\Delta L}\sum_i \frac{1-{\rm FDR}}{w_i\times V_{{\rm max},i}},\end{equation}

\noindent
where $\Delta L$ is the width of the luminosity bin, FDR is the false detection rate, $w_{\rm i}$ is the completeness for the $i$-th galaxy and $V_{\rm max, i}$ is the co-moving volume available to the $i$-th source. For S2CLS sources the false detection rate is (from \citealt{Geach_2017}):

\begin{equation} {\rm log_{10}(FDR)} = 2.67-0.97\times {\rm SNR}. \end{equation}

The errors on the LFs were calculated using the Poissonian approach. The available co-moving volume is the volume between $z_{\rm min}$ and $z_{\rm max}$, where $z_{\rm min}$ is just the lower boundary of a given redshift bin and $z_{\rm max}$ is the maximum redshift at which the $i$-th source would still be visible in a given S2CLS map, or simply the upper boundary of a given redshift bin, whichever is lower. Therefore:

\begin{equation}\label{eq:vmax} V_{{\rm max},i}=\sum_j \frac{\Omega_j}{4\pi}V_{{\rm max}, j},\end{equation}

\noindent
where we sum over all the available S2CLS fields and $\Omega_j$ is the solid angle subtended by the $j$-th field on the sky. The contribution from the $j$-th field to the $V_{{\rm max},i}$, in a given redshift bin, $V_{{\rm max},j}$, is only counted if the maximum redshift at which the $i$-th source would still be visible in that field is higher than the lower boundary of that redshift bin, otherwise $V_{{\rm max},j}$ for that field is simply equal to 0.

Since in practice we are working with 850\,${\rm \mu m}$-detected sources with a mean redshift of $\sim 2.5$ at 850\,${\rm \mu m}$, we decided to initially calculate the LF at a rest-frame wavelength of 250\,${\rm \mu m}$. In order to do so, the luminosity at $850\,{\rm \mu m}\,/\,(1+z)$ is calculated and interpolated to $\lambda_{\rm rest} = 250$\,${\rm \mu m}$ using the average SMG template from \citet{Michalowski_2010}. The range of luminosities and redshifts for which the LFs were calculated were determined from the coverage of the luminosity-redshift plane (grey rectangles in Fig. \ref{fig:lzp}). The luminosity bins used for each field at a given redshift depend on the luminosity lower limit (corresponding to the flux-density detection limit), below which no sources can be detected (solid coloured lines in Fig. \ref{fig:lzp}). Only the luminosity bins with complete luminosity coverage are included in the analysis.

The results, with Poissonian errors, are listed in Table \ref{tab:lf}, and plotted in Fig.\,\ref{fig:lf1}. The coloured solid lines are the best-fitting Schechter functions:

\begin{equation}\label{eq:sch} \Phi_{\rm Sch}(L,z)=\Phi_\star \left(\frac{L}{L_\star}\right)^\alpha exp\left(\frac{-L}{L_\star}\right),\end{equation}

\noindent
with $\Phi_\star$ being the normalisation parameter, $\alpha$ the faint-end slope and $L_\star$ the characteristic luminosity that roughly marks the border between the power-law fit, $(L/L_\star)^\alpha$, and the exponential fit. The black solid lines (almost perfectly aligned with the coloured solid lines) with errors (grey-shaded area) depict the LFs as determined by the maximum-likelihood method (see next subsection). In order to find the faint-end slope, $\alpha$, we utilised the ALMA data (Section \ref{sec:alma}). Since the majority of ALMA sources lie at redshifts $1.5<z<2.5$, it was decided to determine $\alpha$ only for that redshift bin and then use that value in the remaining bins (two faintest points in $1.5<z<2.5$ redshift bin of Fig.\,\ref{fig:lf1}). Fitting the Schechter function to the $1.5<z<2.5$ data yielded a faint-end slope of $\alpha=-0.4$. The remaining Schechter-function parameters were determined by minimising $\chi^2$. In Fig.\,\ref{fig:dlf1} the 68\% and 95\% confidence intervals are shown, corresponding to $\Delta\chi^2=1$ and 4 respectively; the errors on the best-fiting Schechter-function parameters can be determined by projecting the contours onto the relevant axis. The best-fitting values for the Schechter-function parameters are given in Table\,\ref{tab:schparams}, and plotted in Fig.\,\ref{fig:params} as blue circles.

\begin{table}
\begin{normalsize}
\begin{center}
\caption{The best-fitting parameter values for the Schechter functions as determined by the $1/V_{\rm max}$ method. The faint-end slope, $\alpha$, is fixed here at the $z\simeq 2$ value (see text for details).}
\label{tab:schparams}
\setlength{\tabcolsep}{1.0 mm} 
\begin{tabular}{cccc}
\hline
$z$         & $\alpha$  & log($\Phi^\star/{\rm Mpc^{-3} dex^{-1}}$)  & log($L^\star/{\rm W\,Hz^{-1}}$) \\
\hline
$0.5<z<1.5$ & $-0.4$    & $-2.88^{+0.33}_{-0.27}$      & $25.20^{+0.08}_{-0.09}$ \\
$1.5<z<2.5$ & $-0.4$    & $-3.03^{+0.05}_{-0.10}$      & $25.40^{+0.03}_{-0.03}$ \\
$2.5<z<3.5$ & $-0.4$    & $-3.73^{+0.13}_{-0.16}$      & $25.63^{+0.05}_{-0.05}$ \\
$3.5<z<4.5$ & $-0.4$    & $-4.59^{+0.28}_{-0.31}$      & $25.84^{+0.16}_{-0.14}$ \\
\hline
\end{tabular}
\end{center}
\end{normalsize}
\end{table}

\subsection{Maximum-likelihood method}
\label{sec:meth2}

The likelihood function used here \citep{Marshall_1983} is defined as a product of the probabilities of observing exactly one source in $dzdL$ at the position of the $i$-th galaxy ($z_i, L_i$) for $N$ galaxies in our sample and of the probabilities of observing zero sources in all the other differential elements in the luminosity-redshift plane. Using Poisson probabilities, the likelihood is:

\begin{equation}\label{eq:likel} \mathcal{L}=\prod_i^N \lambda_i e^{-\lambda_i}\,\prod_j e^{-\lambda_j}, \end{equation}

\noindent
where $j$ runs over all differential elements in which no sources were observed, and $\lambda$ is the expected number of galaxies in $dzdL$ at $z,L$:

\begin{equation}\label{eq:lambda} \lambda=\Phi(z,L)\Omega(z,L)\frac{dV}{dz}dzdL, \end{equation}

\noindent
with $\Omega$ being the fractional area of the sky in which a galaxy with a given $z$ and $L$ can be detected in our fields. Since for large $N$ the test statistic $-2\,{\rm ln}(\mathcal{L})$ will be $\chi^2$ distributed we define:

\begin{multline}\label{eq:s} S=-2\,{\rm ln}(\mathcal{L})=-2\sum_i^N{\rm ln}[\Phi(z_i,L_i)]+\\
2\,\iint \Phi(z,L)\Omega(z,L)\frac{dV}{dz}dzdL,
\end{multline}

\noindent
where we dropped terms independent of the model parameters. This step transforms both exponents in Equation \ref{eq:likel} into the integral, which represents the model $-$ the expected number of sources within the integral limits. This means that, as compared to the $1/V_{\rm max}$ method, this technique can analyse the entire luminosity-redshift plane (including its empty patches) and therefore give a result with higher statistical significance. In addition, it does not bin the data, and also allows a user to define the parameters of a luminosity function, $\Phi(z,L)$, to be continuous functions of redshift, which can be then determined by minimising $S$.

For consistency, we chose $\Phi(z,L)$ to have the form of a Schechter function (Equation \ref{eq:sch}). As before, the faint-end slope, $\alpha$, was fixed at the $1.5<z<2.5$ value of $-0.4$. As we sought an acceptable description of the data, we explored various functional forms for the redshift dependence of $\Phi_\star$ and log($L_\star$), and in the end found that we could adopt a simple normalised Gaussian function\footnote{The normalised Gaussian was chosen here instead of the linear function to ensure that $\Phi_\star$ does not become negative at high-$z$.} and a linear function of redshift respectively:

\begin{equation}
\begin{aligned}
\Phi_\star(z\,|\,A,\sigma,\mu) &= \frac{A}{\sigma\sqrt{2\pi}}e^{-\frac{(z-\mu)^2}{2\sigma^2}}\\
{\rm log}(L_\star(z\,|\,a,b)) &= az+b.
\end{aligned}
\label{eq:params}
\end{equation}

\begin{table}
\begin{footnotesize}
\begin{center}
\caption{Best-fit values for the continuous functions of the Schechter function parameters from Equation \ref{eq:params}, with $1\sigma$ errors based on the maximum-likelihood ratios and corresponding to $\Delta \chi^2=1$.}
\label{tab:params}
\setlength{\tabcolsep}{3.5 mm} 
\begin{tabular}{cc}
\hline
\hline
Parameter & Value \\
\hline
$A$       & $2.40^{+0.60}_{-0.48}\times 10^{-3}$  \\
$\sigma$  & $1.04^{+0.06}_{-0.05}$  \\
$\mu$     & $1.28^{+0.20}_{-0.25}$  \\
$a$       & $0.19^{+0.02}_{-0.02}$  \\
$b$       & $25.03^{+0.06}_{-0.05}$ \\
\hline
\end{tabular}
\end{center}
\end{footnotesize}
\end{table}

\begin{figure*}
\begin{center}
\includegraphics[scale=0.8]{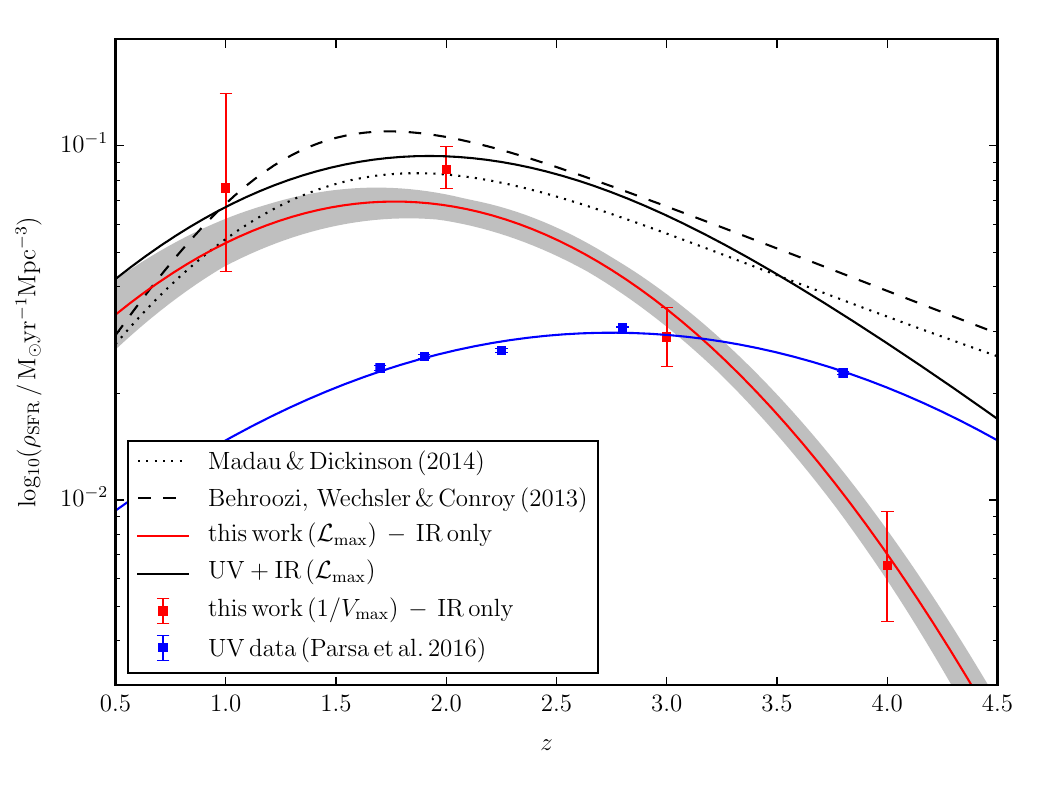}
\end{center}
\caption{The star-formation rate density, $\rho_{\rm SFR}$, as a function of redshift. The black dotted and dashed lines represent the recent parametric descriptions of the redshift evolution of the $\rho_{\rm SFR}$ provided by \citet{Madau_2014} and \citet{Behroozi_2013} respectively (both converted to a Chabrier IMF). The red filled squares show our estimates of the IR SFRDs as determined using the $1/V_{\rm max}$ method, converting from IR luminosity density to SFRD using the conversion factor of $\mathcal{K}_{\rm IR}=4.5 \times 10^{-44}\, {\rm M_\odot\, year^{-1}\, erg^{-1}\, s}$ given in \citet{Kennicutt_1998}, multiplied by a factor of 0.63 to convert from a Salpeter to a Chabrier IMF. The red solid line (with the $1\sigma$ uncertainty depicted by the grey-shaded area) shows our analogous estimate of the redshift evolution of the IR SFRD, as determined from the luminosity-weighted integration of the LF derived through the maximum-likelihood method (with the functional form given in Equation \ref{eq:sfrdir}). The blue squares are the UV SFRD estimates based on the results of  \citet{Parsa_2016}, converting from the rest-frame UV (1500\AA) luminosity to UV-visible SFR using the factor  $\mathcal{K}_{\rm UV}=1.3 \times 10^{-28}\, {\rm M_\odot\, year^{-1}\, erg^{-1}\, s\, Hz}$ from \citet{Madau_2014} (again with a further correction factor of $\times$0.63 to convert to a Chabrier IMF). The blue solid line depicts a best-fitting function to these UV-derived results given by Equation \ref{eq:sfrduv}. The black solid line shows a functional form of the total $\rho_{\rm SFR}$ given by adding the UV and IR results.}
\label{fig:sfrd}
\end{figure*}

The redshift limits of the integral in Equation \ref{eq:s} are the same as in the $1/V_{\rm max}$ method, ${\rm z}=0.5-4.5$. The luminosity limits were chosen in order to cover the whole useful luminosity range (including the ALMA data), ${\rm log_{10}}(L_{250}\,/\,{\rm W\,Hz^{-1}})=24.6-26.6$. Minimising $S$ gave the best-fitting parameter values, as summarized in Table\,\ref{tab:params}, which were then used to determine the redshift evolution of the best-fitting Schechter-function model parameters, $\Phi_\star$ and $L_\star$, as depicted in Fig.\,\ref{fig:params} by the blue solid lines. Since, for the large number of sources we have here, $S=-2\,{\rm ln}(\mathcal{L})$ is $\chi^2$ distributed, the $1\sigma$ errors (grey area in Fig.\,\ref{fig:params}) are based on the likelihood ratios and correspond to $\Delta \chi^2=1$. For comparison the best-fitting values (with $1\sigma$ errors) for four redshift bins, calculated using the $1/V_{\rm max}$ method, are also plotted here as blue points.

With the redshift evolution of the Schechter-function parameters as described in Equation \ref{eq:params}, it is straightforward to calculate the luminosity function at any redshift between $z = 0.5$ and $z = 4.5$. We therefore plot the LFs determined using the maximum-likelihood method alongside the results of the previous subsection in Fig.\,\ref{fig:lf1} as black solid lines. The $1\sigma$ errors (grey area) are again based on the maximum-likelihood ratios and correspond to $\Delta \chi^2=1$.

\section{Star Formation Rate Density}
\label{sec:sfrd}

\begin{figure*}
\begin{center}
\includegraphics[scale=0.775]{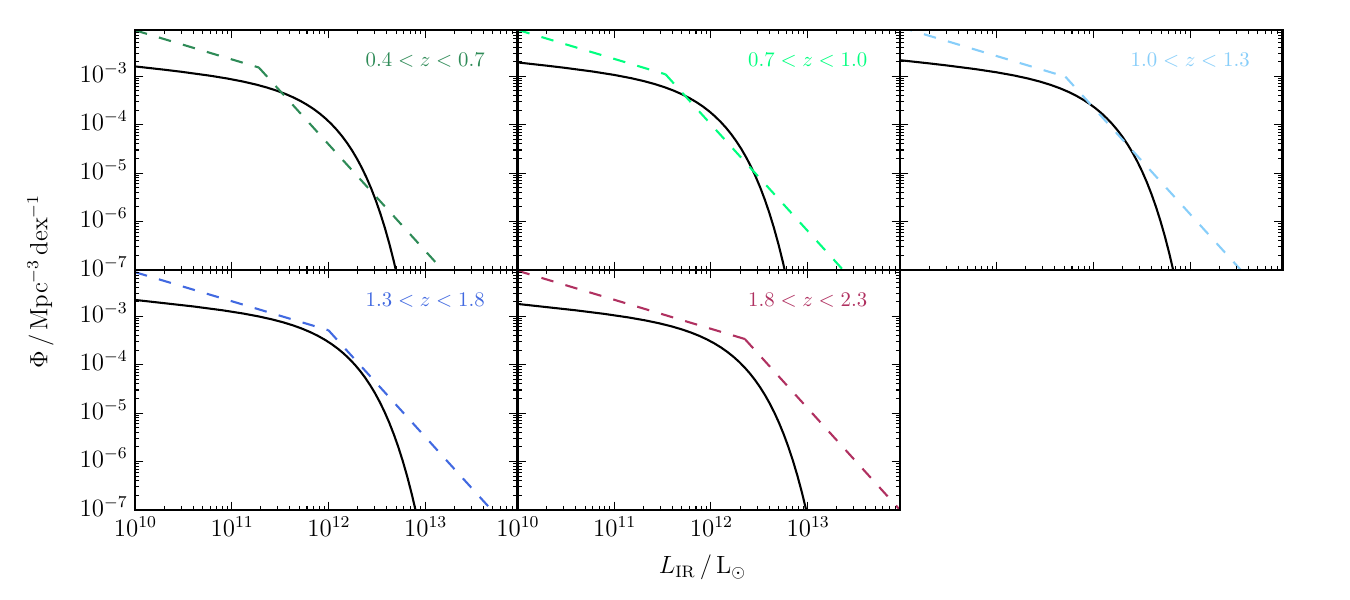}
\end{center}
\caption{A comparison of the IR LFs found in this work with those of \citet{Magnelli_2013}, depicted as black solid lines and coloured dashed lines respectively. The solid lines are our IR LFs plotted using Equation \ref{eq:sch} (scaled to IR luminosity using the average SMG SED of \citealt{Michalowski_2010}), with the parameters found at the median redshift of each bin using Equation \ref{eq:params}. It can be seen that our LFs and those of \citet{Magnelli_2013} generally agree reasonably well near the break luminosity, but differ substantially at both faint and bright luminosities in every redshift bin. As discussed further in the text, much of this apparent difference is driven by the different parameterizations adopted. At the bright end, the \citet{Magnelli_2013} sample in fact contains only one object with an estimated  $L_{\rm} > 10^{13}\,{\rm L_{\odot}}$, and their adopted bright-end fixed power-law fit results in a severe over-prediction (by a factor $\simeq 15-20$) of the number of bright sub-mm sources actually found in the degree-scale SCUBA-2 surveys. At the faint end, \citet{Magnelli_2013} fixed all their LFs to have a faint-end power-law slope of $-0.6$, whereas we have adopted a slightly shallower slope of $-0.4$, derived by incorporating the ALMA HUDF results at $z \simeq 2$.}
\label{fig:lfcomM}
\end{figure*}

Having constructed the rest-frame 250-${\rm \mu m}$ LFs, it is now possible to establish the redshift evolution of the star formation rate density ($\rho_{\rm SFR}$). This is achieved by first integrating the LFs weighted by the total IR luminosity ($8-1000\,{\rm \mu m}$); for consistency this was performed again assuming the SMG SED of \citet{Michalowski_2010} (which, although peaking at a wavelength consistent with a temperature of 35\,K, yields a bolometric luminosity $\simeq 2$ times greater than a simple 35\,K modified blackbody template), and the LFs were integrated down to a lower luminosity limit of $0.01\times L^\star$. The resulting inferred total IR luminosity density, $\rho_{\rm IR}$ was then converted into dust-obscured star-formation rate density, $\rho_{\rm SFR}$, at each redshift using the scaling factor of $\mathcal{K}_{\rm IR}= 4.5 \times 10^{-44}\, {\rm M_\odot\, year^{-1}\, erg^{-1}\, s}$ from \citet{Kennicutt_1998}, with an additional multiplicative factor of 0.63 to convert from a Salpeter to a Chabrier IMF.

The results are shown in Fig.\,\ref{fig:sfrd}. The red filled squares depict the values of $\rho_{\rm SFR}$ derived from the four LFs shown in Fig.\,\ref{fig:lf1}, established using the $1/V_{\rm max}$ method. The red solid line shows the evolution of $\rho_{\rm SFR}$ as determined from the continuous form of the redshift evolution of the LF found using the maximum-likelihood method, with the grey area showing the $1\sigma$ errors. Due to the fact we have chosen to parametrize the Schechter function normalisation parameter, $\Phi^\star$ as a Gaussian function , and have chosen a simple linear dependence of redshift for ${\rm log}(L_{\star})$ (see Equation \ref{eq:params}), the redshift evolution of $\rho_{\rm SFR}$, depicted as a red solid line, is also a Gaussian:

\begin{equation}\label{eq:sfrdir} \rho_{\rm SFR_{IR}}(z) = \frac{0.18}{1.04\sqrt{2\pi}}e^{-\frac{(z-1.77)^2}{2\times 1.04^2}}. \end{equation}

The blue squares give the values of UV-visible $\rho_{\rm SFR}$ derived from the \citet{Parsa_2016} $\rho_{\rm UV}$ results (integrated down to $M_{\rm UV} = -10$) by converting to UV-visible $\rho_{\rm SFR}$ using the \citet{Madau_2014} scaling of $\mathcal{K}_{\rm UV}=1.3 \times 10^{-28}\, {\rm M_\odot\, year^{-1}\, erg^{-1}\, s\, Hz}$, again also multiplied by a factor of 0.63 to convert to a Chabrier IMF. The blue solid line is a best-fitting Gaussian function to the UV results:

\begin{equation}\label{eq:sfrduv} \rho_{\rm SFR_{UV}}(z) = \frac{0.11}{1.48\sqrt{2\pi}}e^{-\frac{(z-2.75)^2}{2\times 1.48^2}}. \end{equation}

The black solid line shows the total $\rho_{\rm SFR}$ calculated by simply adding the IR and UV estimates (Equations \ref{eq:sfrdir} and \ref{eq:sfrduv} respectively). The black dotted and dashed lines depict the alternative functional forms of the redshift evolution of total $\rho_{\rm SFR}$ provided by \citet{Madau_2014} and \citet{Behroozi_2013} respectively (after conversion to a Chabrier IMF).

\section{Discussion}
\label{sec:disc}

\subsection{Comparison of luminosity functions}
\label{sec:lf_compare}

\begin{figure*}
\begin{center}
\includegraphics[scale=0.775]{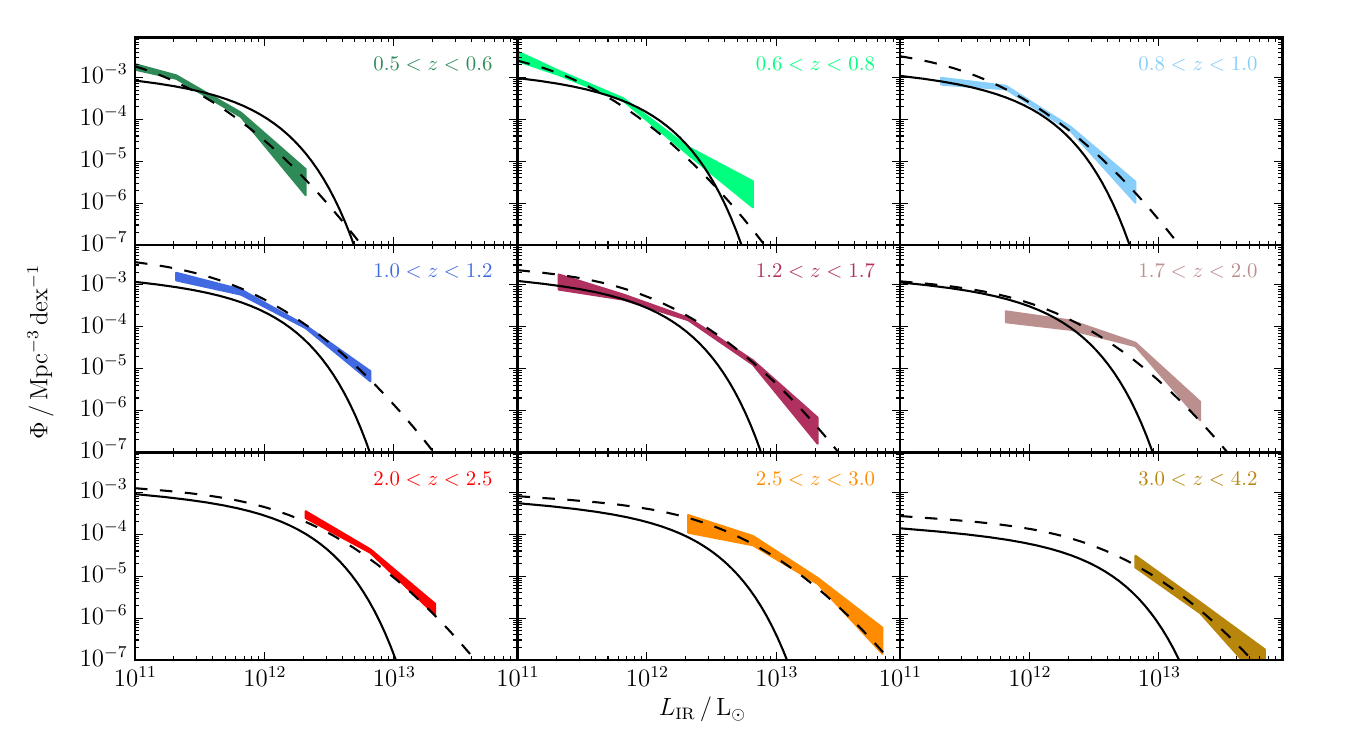}
\end{center}
\caption{A comparison of the IR LFs found in this work with those of \citet{Gruppioni_2013}, depicted as black solid lines and coloured areas respectively. The dashed black lines are the best-fitting modified Schechter functions fitted by \citet{Gruppioni_2013} to their binned LF data. The faint-end slope assumed by \citet{Gruppioni_2013} was fixed at $-$0.2, whereas in the present study it was fixed at the measured $z\simeq 2$ value of $-0.4$. In most redshift bins, there is reasonable agreement at the faint end of the LF. However, certainly at $z > 1$, the \citet{Gruppioni_2013} LFs are much higher (or the sources are much brighter) at the bright end. This difference manifests itself clearly in the inferred values of IR $\rho_{\rm SFR}$, where the \citet{Gruppioni_2013} LFs yield significantly higher values than the present study, especially at $z \simeq 3$ (Fig.\,\ref{fig:sfrdc}). In addition, the predicted cumulative number counts at 850\,${\rm \mu m}$ are very different (see Fig.\,\ref{fig:nc} and Table \ref{tab:nc}); the \citet{Gruppioni_2013} LFs predict $>300$ 850-${\rm \mu m}$ sources with flux densities $S_{850} > 10$\,mJy in the maps used in this work, whereas in fact only 17 such objects are actually detected.}
\label{fig:lfcomG}
\end{figure*}

It is interesting to compare our inferred total IR LFs with those that have been derived from the {\it Herschel} surveys. First, in Fig. \ref{fig:lfcomM}, we compare our LFs with those produced by \citet{Magnelli_2013} from the deepest PACS surveys in the GOODS fields. \citet{Magnelli_2013} utilised the 70, 100 and 160-$\mu$m imaging produced by the PACS Evolutionary Probe (PEP; \citealt{Lutz_2011}) and GOODS-{\it Herschel} (GOODS-H; \citealt{Elbaz_2011}) programmes. They performed blind PACS source extraction, but also created a source catalogue using positional priors, with the positions of {\it Spitzer} IRAC 3.6-$\mu$m sources used to extract sources from the {\it Spitzer} MIPS 24-$\mu$m imaging, which were then in turn used as positional priors for source extraction in the PACS maps. The latter (MIPS-PACS) source catalogues were then cross-matched with the shorter-wavelength GOODS catalogues (optical+near-infrared) to provide the required photometric redshift information, and the total infrared (8--1000-$\mu$m) luminosities were inferred by fitting the 70, 100 and 160-$\mu$m photometry with the SED template library of \citet{Dale_2002}. \citet{Magnelli_2013} used this information to construct LFs in six redshift bins using the $1/V_{\rm max}$ method, and fitted the binned LF data with a double power-law function as used previously by \citet{Magnelli_2009, Magnelli_2011}. Because these LFs are based on the PACS data, which do not extend to wavelengths longer than 160\,$\mu$m, \citet{Magnelli_2013} confined their analysis to redshifts $z < 2.3$.

Our results are compared with the \citet{Magnelli_2013} LFs in Fig.\ref{fig:lfcomM}, using the five redshift bins adopted by \citet{Magnelli_2013}. The coloured dashed lines show the \citet{Magnelli_2013} LFs, while the solid line in each panel shows our LF at the medium redshift of each bin. With the possible exception of the highest redshift bin, the two sets of LFs agree fairly well near the break luminosity. As a result, it transpires that, as shown in the next section, the derived IR luminosities densities are not very different (albeit the \citet{Magnelli_2013} results are systematically higher). Nevertheless, it is clear that the LFs diverge at both lower and higher luminosities, with the \citet{Magnelli_2013} LFs indicating larger numbers of both faint and bright sources. However, to some extent this difference is exaggerated by the different parameterisations adopted. 

Specifically, the difference at the bright end is due, at least in part, to the fact that \citet{Magnelli_2013} adopted a double power-law function, while we have fitted a Schechter function (with an exponential cutoff at bright luminosities). In reality, the \citet{Magnelli_2013} sample contains only one object with an estimated  $L_{\rm} > 10^{13}\,{\rm L_{\odot}}$ and only $\simeq 6$ sources with $L_{\rm} > 5 \times 10^{12}\,{\rm L_{\odot}}$ across the entire redshift range (their figure 8). As a consequence, the bright end of the \citet{Magnelli_2013} LFs is fairly unconstrained, and much of the difference seen in Fig. \ref{fig:lfcomM} is actually driven by their adoption of a fixed bright-end power-law slope. In fact, as discussed below, integrated over the redshift range $0.5 < z < 2.3$ the \citet{Magnelli_2013} LFs predict $\simeq 20$ times as many bright 850-$\mu$m sources in the wide-area SCUBA-2 sources than are actually observed, and so their LFs are clearly much too high at the bright end.

\begin{figure*}
\begin{center}
\includegraphics[scale=0.58]{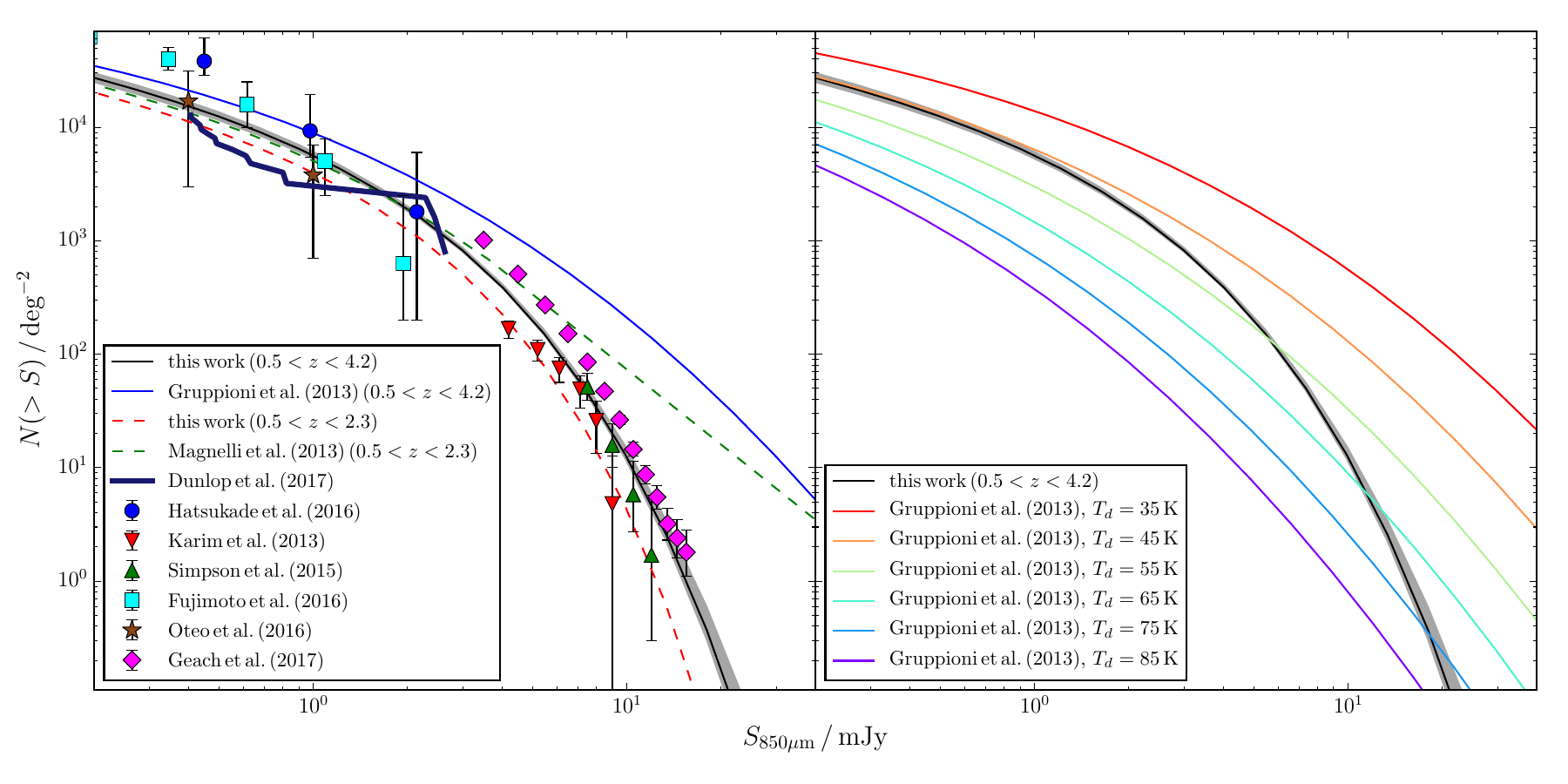}
\end{center}
\caption{Cumulative number counts as a function of 850-${\rm \mu m}$ flux density. In the {\bf left-hand} panel we show a range of published cumulative counts from the SCUBA-2 and ALMA literature \citep{Karim_2013, Simpson_2015, Hatsukade_2016, Fujimoto_2016, Oteo_2016, Dunlop_2017, Geach_2017}, along with the predicted number counts calculated from our new IR LFs (presented and discussed in Section \ref{sec:lf_compare}), after integration over the redshift range of $0.5<z<4.5$ (black solid line, with the $1\sigma$ uncertainty shown by the grey-shaded area), and after integration over the redshift range $0.5<z<2.3$ (red dashed line). For comparison, the 850-$\mu$m number counts predicted by integrating the \citet{Gruppioni_2013} evolving LF over $0.5<z<4.5$ are plotted as the blue solid line, while the counts predicted by integrating the \citet{Magnelli_2013} LFs  over $0.5<z<2.3$ are indicated by the green dashed line. These predictions have all been made assuming the SED of \citet{Michalowski_2010} as used throughout this study. Unsurprisingly, since our study is based primarily on 850-$\mu$m data, the number counts predicted by our new IR LFs agree well with the data, whereas the {\it Herschel}-based LFs over predict the counts at the bright end by over an order-of-magnitude (as expected, given the dramatic differences at the bright end of the LFs, as shown in Fig. \ref{fig:lfcomM} and Fig. \ref{fig:lfcomG}). These over-predictions are quantified in Table \ref{tab:nc}, for 850-${\rm \mu m}$ sources with flux densities higher than  in the combined area of all the S2CLS fields used in this work ($\simeq 1.5$ deg$^2$). The actual number of $S_{850} > 10$\,mJy sources in this area in the real data is 17 sources in the redshift range $0.5<z<4.5$, or  4 sources for $0.5<z<2.3$. In the {\bf right-hand} panel we explore the extent to which such discrepancies can potentially be solved by varying the adopted SED template. To illustrate this we show the effect of converting the \citet{Gruppioni_2013} LFs into predicted 850-$\mu$m number counts for adopted modified black-body SEDs of increasing temperature. As expected, things agree well at the faint end for $T \simeq 45$\,K, but a change to characteristic temperatures of $T \simeq 80$\,K is required to suppress the predicted counts at the bright end. There is no evidence to support such a rapid change in typical SED effective temperature over such a short range in 850-$\mu$m flux density of IR luminosity.}
\label{fig:nc}
\end{figure*}

At the faint end, \citet{Magnelli_2013} again adopted a fixed power-law slope, with $\phi \propto L^{-0.6}$. This was based on the local value derived by \citet{Sanders_2003}, and simply fixed at higher redshifts because the PACS data did not really enable meaningful constraints to be placed on the faint-end slope at $z > 0.5$. Thus, in Fig. \ref{fig:lfcomM}, our LFs diverge from those of \citet{Magnelli_2013} towards fainter luminosities because we have adopted a slightly shallower faint-end slope ($\phi~\propto~L^{-0.4}$), based on the ALMA measurements at $1.5 < z < 2.5$. As discussed in more detail next, some other {\it Herschel}-based estimates of the faint end of the LF in fact use an even flatter faint-end slope ($\phi \propto L^{-0.2}$; \citealt{Gruppioni_2013}) and, as shown in figure 11 of \citet{Magnelli_2013}, even using the {\it Spitzer} MIPS data to extend estimates of the LF to fainter luminosities cannot really distinguish between a faint-end power-law slope of $-0.6$ or $-0.2$. It is therefore both interesting and reassuring that our own measured value of the faint-end slope ($-0.4$) lies midway between the values adopted in the {\it Herschel} studies.

Next, in Fig. \ref{fig:lfcomG}, we compare our LFs with those produced by \citet{Gruppioni_2013}, which were based on PACS far-IR sources extracted from the {\it Herschel} PEP survey data at 70, 100 and 160\,${\rm \mu m}$ within the COSMOS, ECDFS, GOODS-N and GOODS-S fields (covering a total area $\simeq 3.3$\,deg$^2$). As well as covering a larger area than \citet{Magnelli_2013} (albeit of course to shallower depths), \citet{Gruppioni_2013} extended the SED fitting to include the 250, 350 and 500-${\rm \mu m}$ imaging provided by the {\it Herschel} SPIRE imaging in the same fields (by the HerMES survey; \citealt{Oliver_2012}), and attempted to extend their LF analysis out to redshifts $z \simeq 4$. \citet{Gruppioni_2013} detected 373, 7176 and 7376 sources at 70, 100 and 160\,${\rm \mu m}$, respectively, used cross-matching (via 24-$\mu$m MIPS imaging) with the shorter-wavelength (optical+near-IR) data in the fields to provide redshift information, and used SED fitting with a range of templates to the PACS+SPIRE photometry to derive total IR luminosities. Again, \citet{Gruppioni_2013} constructed the LFs using the $1/V_{\rm max}$ method, but in their study they fitted the binned LF data with a modified Schechter function.

Our results are compared with the \citet{Gruppioni_2013} LFs in Fig.\,\ref{fig:lfcomG}, using the nine redshift bins adopted by \citet{Gruppioni_2013}. The colour-shaded areas show the \citet{Gruppioni_2013} results, with the dashed lines representing their best-fitting modified Schechter functions. The solid black lines show our own estimates of the LF at these redshifts. While at the faint-end the LFs are in fairly good agreement (as mentioned above, \citet{Gruppioni_2013} in fact adopt a slightly shallower faint-end slope), at the bright-end there is again significant disagreement. Moreover, at high redshifts the \citet{Gruppioni_2013} LFs are significantly higher around the break luminosity, resulting in substantially larger values of IR luminosity density at $z \simeq 3$ (see next section).

\begin{table}
\begin{footnotesize}
\begin{center}
\caption{The predicted number of  850-${\rm \mu m}$ sources with flux densities $S_{850} > 10$\,mJy in the combined area of all the fields used in this study ($\simeq 1.5$ deg$^2$) produced by integrating the three IR LFs discussed in Section \ref{sec:lf_compare} and shown in Fig. \ref{fig:lfcomM} and Fig. \ref{fig:lfcomG}. As discussed in the caption to Fig. \ref{fig:nc}, and indicated here, these predictions were based on integration over the redshift range $0.5<z<4.5$ or $0.5 < z < 2.3$ as appropriate. The actual number of sources in our sample with $S_{850} > 10$\,mJy is 17 in the redshift range $0.5<z<4.5$, of which four have $0.5<z<2.3$. The predictions are based on the \citet{Michalowski_2010} average sub-mm source SED. Clearly, the {\it Herschel}-derived LFs wildly over-predict the actual number of bright 850-$\mu$m sources.} 
\label{tab:nc}
\setlength{\tabcolsep}{3.5 mm} 
\begin{tabular}{lcc}
\hline
\hline
Study                  & z range     & $N(>10\,{\rm mJy})$ \\
                       &             & $/\,1.5\,{\rm deg}^{-2}$ \\
\hline
This Work              & $0.5<z<4.5$ & 20  \\
\citet{Gruppioni_2013} & $0.5<z<4.5$ & 315 \\
This Work              & $0.5<z<2.3$ & 8   \\
\citet{Magnelli_2013}  & $0.5<z<2.3$ & 142 \\
\hline
\end{tabular}
\end{center}
\end{footnotesize}
\end{table}

\begin{figure*}
\begin{center}
\includegraphics[scale=0.775]{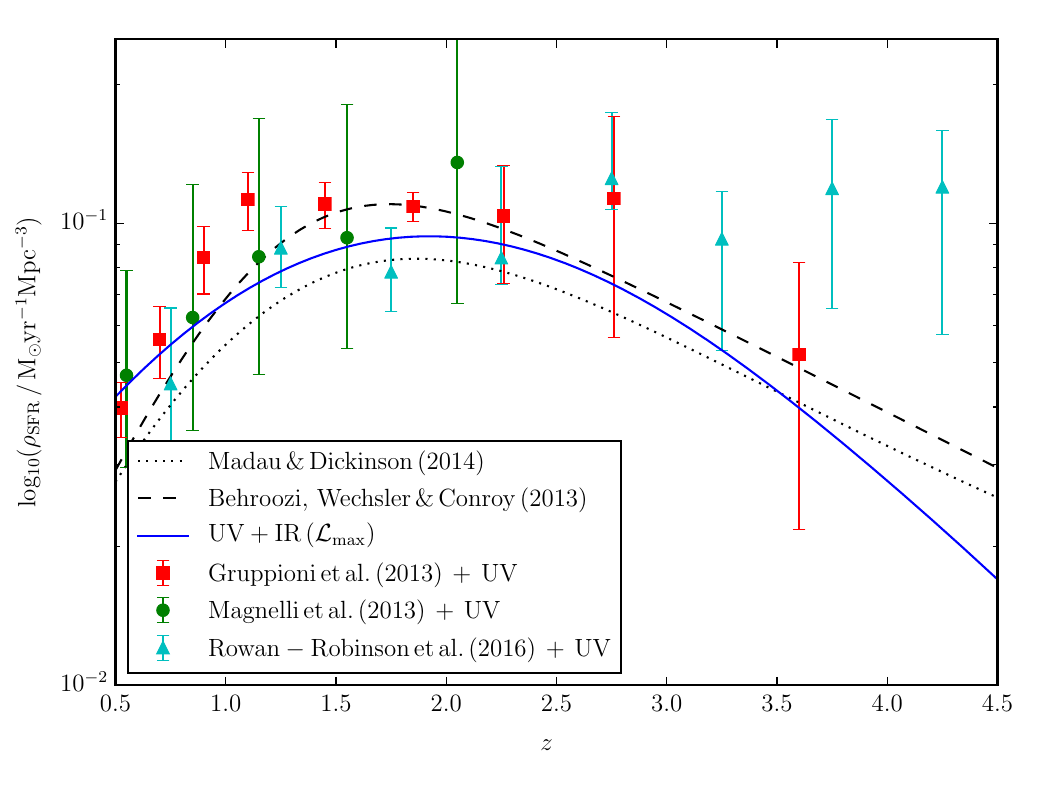}
\end{center}
\caption{Total $\rho_{\rm SFR}$ as a function of redshift. The blue solid line shows the results of this work (see Fig. \ref{fig:sfrd}), while the data points show the estimates derived from the {\it Herschel}-based work of \citet{Magnelli_2013} (green circles),  \citet{Gruppioni_2013} (red squares) and \citet{Rowan_2016} (magenta triangles) after addition of the UV estimates of unobscured $\rho_{\rm SFR}$ from \citet{Parsa_2016}. The black dotted and dashed lines represent the recent parametric descriptions of the redshift evolution of $\rho_{\rm SFR}$ found by \citet{Madau_2014} and \citet{Behroozi_2013} respectively, adopting a Chabrier IMF.}
\label{fig:sfrdc}
\end{figure*}

One way to quantify how much our new IR LFs differ from those produced from the aforementioned {\it Herschel}-based studies is to calculate how many bright sub-mm sources each evolving LF predicts should be present in the SC2LS survey data utilised here. We show the results of this in Fig. \ref{fig:nc}, and quantify the differences at flux densities $S_{850} > 10$\,mJy in Table \ref{tab:nc}. Along with a range of data derived from the SCUBA-2 and ALMA-based literature, the solid black and blue lines in the left-hand panel of Fig.\,\ref{fig:nc} show the predicted cumulative 850-$\mu$m number counts derived from our new LFs and from those produced by \citet{Gruppioni_2013} (see Fig. \ref{fig:lfcomG}) after integration over the redshift range $0.5<z<4.5$. The dashed red and green lines show the corresponding predictions from our own LFs and those produced by \citet{Magnelli_2013} (see Fig. \ref{fig:lfcomM}) after integration over the redshift range $0.5<z<2.3$. Unsurprisingly, since our study is based primarily on 850-$\mu$m data, the number counts predicted by our new IR LFs agree well with the observed counts, whereas the {\it Herschel}-based LFs over predict the counts at the bright end by over an order-of-magnitude (as expected, given the dramatic differences at the bright end of the LFs, as shown in Fig. \ref{fig:lfcomM} and Fig. \ref{fig:lfcomG}).

The extent of this difference is quantified further in Table \ref{tab:nc}, where we tabulate how many 850-${\rm \mu m}$ sources with flux densities $S_{850} >10$\,mJy in the combined area of all the S2CLS fields used here (1.5\,deg$^2$) a given LF predicts after integration over the appropriate redshift range. The actual detected number of 850-${\rm \mu m}$ sources with $S_{850} > 10$\,mJy  is 17 over the redshift range $0.5<z<4.5$ (consistent with our predicted 20) of which 4 have $0.5<z<2.3$ (consistent with our predicted 8). By contrast the \citet{Gruppioni_2013} LFs predict $>300$ such galaxies at $0.5<z<4.5$ (higher by a factor of $\simeq 15-20$), while the \citet{Magnelli_2013} LFs predict 140 such sources at $0.5<z<2.3$ (again high by a factor of $15-20$). 

As used consistently throughout this study, conversion between total IR luminosity, and observed 850-$\mu$m flux density was performed using the average sub-mm source SED of \citet{Michalowski_2010}. In the right-hand panel of Fig.\,\ref{fig:nc} we explore what sort of change in the adopted template would be required to bring the bright sub-mm number-count predictions of the {\it Herschel}-derived IR LFs into agreement with reality. Here we show the effect of converting the \citet{Gruppioni_2013} LFs into predicted 850-$\mu$m number counts for adopted modified black-body SEDs of increasing temperature. As expected, things agree well at the faint end for $T \simeq 45$\,K, but a change to characteristic temperatures of $T \simeq 80$\,K is required at the bright end to adequately suppress the predicted number counts. There is no evidence to support such a rapid change in typical SED temperature over such a short range in 850-$\mu$m flux density or IR luminosity. Indeed, while several studies have reported a correlation between IR luminosity and the effective temperature of the typical galaxy SED, the derived slope of this correlation is relatively modest, with average dust temperature reported to rise from $T \simeq 30$\,K for $L_{\rm IR} \simeq 10^{11}\,{\rm L_{\odot}}$ to $T \simeq 40$\,K for $L_{\rm IR} \simeq 10^{13}\,{\rm L_{\odot}}$ \citep{Amblard_2010, Hwang_2010, Smith_2012}. 

We conclude that the steep drop-off at the bright end of the sub-mm number counts does simply reflect a near exponential decline at the bright end of the IR LF, and the {\it Herschel} results have been contaminated and biased high at the bright end by a mix of blending issues, source identification (and hence redshift) errors, and possibly also the adoption of a double power-law LF (see also \citealt{Bethermin_2017, Liu_2017}).

\begin{figure*}
\begin{center}
\includegraphics[scale=0.7]{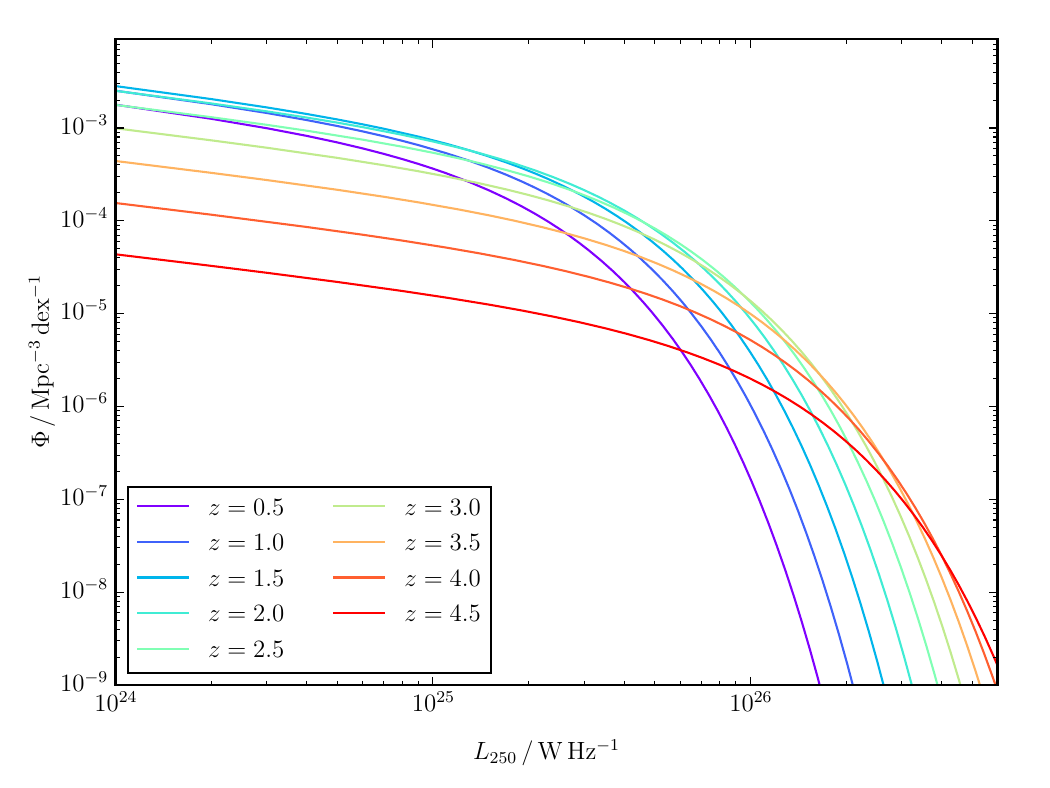}
\end{center}
\caption{Our derived rest-frame 250-${\rm \mu m}$ luminosity functions (LFs) for a range of redshifts, determined from the maximum-likelihood method (plotted using Equations \ref{eq:sch} and \ref{eq:params}, using the best-fit parameter values listed in Table \ref{tab:params}). The combined impact of rising-then-falling density evolution and positive luminosity evolution with redshift is clear. The result is that, as with AGN, many of the most luminous sources are to be found at high redshifts even though overall IR luminosity density is in decline beyond $z \simeq 2$. The implications for the expected redshift distribution of 850-$\mu$m sources as a function of flux density are illustrated in Fig. \ref{fig:zcomp}.}
\label{fig:lf2}
\end{figure*}

\subsection{Comparison of star-formation rate density}

In Fig. \ref{fig:sfrdc} we compare the redshift evolution of cosmic star-formation density derived here with the evolution inferred from the {\it Herschel}-based studies undertaken by \citet{Magnelli_2013}, \citet{Gruppioni_2013} and \citet{Rowan_2016}. All results have been converted to a Chabrier IMF, and the values of obscured $\rho_{\rm SFR}$ derived from the sub-mm/far-IR studies have been added to the UV-derived $\rho_{\rm SFR}$ from \citet{Parsa_2016}, so that all values plotted here represent {\it total} $\rho_{\rm SFR}$.

To derive the \citet{Magnelli_2013} results (green points in Fig. \ref{fig:sfrdc}) we performed the luminosity-weighted integral of their IR LFs (shown in Fig. \ref{fig:lfcomM}) from a lower limit of $0.01\times L^\star$, where $L^\star$ is the characteristic luminosity found for our sample (Table \ref{tab:schparams}). We then derived $\rho_{\rm SFR}$ using the conversion factor given by \citet{Madau_2014}, before applying the IMF correction factor of 0.63 and adding the UV-based results. To derive the corresponding values from \citet{Gruppioni_2013}, we simply converted their own IR luminosity density estimates using the same procedure. \citet{Rowan_2016} attempted to extend the exploitation of the {\it Herschel} HerMES survey to the highest redshifts via SPIRE-based source selection (over an area of $\simeq 20$\,deg$^2$), including sources detected only at 500\,$\mu$m. We have adopted their derived estimates of obscured $\rho_{\rm SFR}$, converted to a Chabrier (2003) IMF, and again added the UV-based estimates to produce the (magenta) data points shown in Fig. \ref{fig:sfrdc}. 

Despite the differences in the LFs discussed extensively above, with the exception of the rather high \citet{Gruppioni_2013} results around $z \simeq 1$, all three {\it Herschel}-based studies yield estimates of $\rho_{\rm SFR}$ consistent with those derived here (as indicated by the blue curve) up to redshifts $z \simeq 2.2$, albeit the uncertainties in the {\it Herschel}-derived  estimates are generally very large. At higher redshifts the uncertainties in the {\it Herschel}-based estimates start to approach 0.5 dex, but even so the smooth decline in $\rho_{\rm SFR}$ beyond $z \simeq 2$ seen in the present study (and consistent with that reported by \citealt{Behroozi_2013, Madau_2014, Dunlop_2017, Bourne_2017, Liu_2017}) is inconsistent with the high values reported by \citet{Rowan_2016}. We suggest that the high values derived by \citet{Rowan_2016} may reflect problems in source identification and redshift estimation stemming from the large-beam long-wavelength SPIRE data, as well as potential blending issues, and extrapolation of $\rho_{\rm SFR}$ from the most extreme sources. Interestingly, the \citet{Rowan_2016} results seem somewhat low near the peak of activity at $z \simeq 2$. This may suggest that some subset of numerous far-IR sources which should lie at $z \simeq 2$ have been erroneously placed at higher redshifts, where the resulting boost in inferred luminosity coupled with the smaller cosmological volumes will inevitably result in artificially high estimates of $\rho_{\rm SFR}$.

\begin{figure*}
\begin{center}
\includegraphics[scale=0.7]{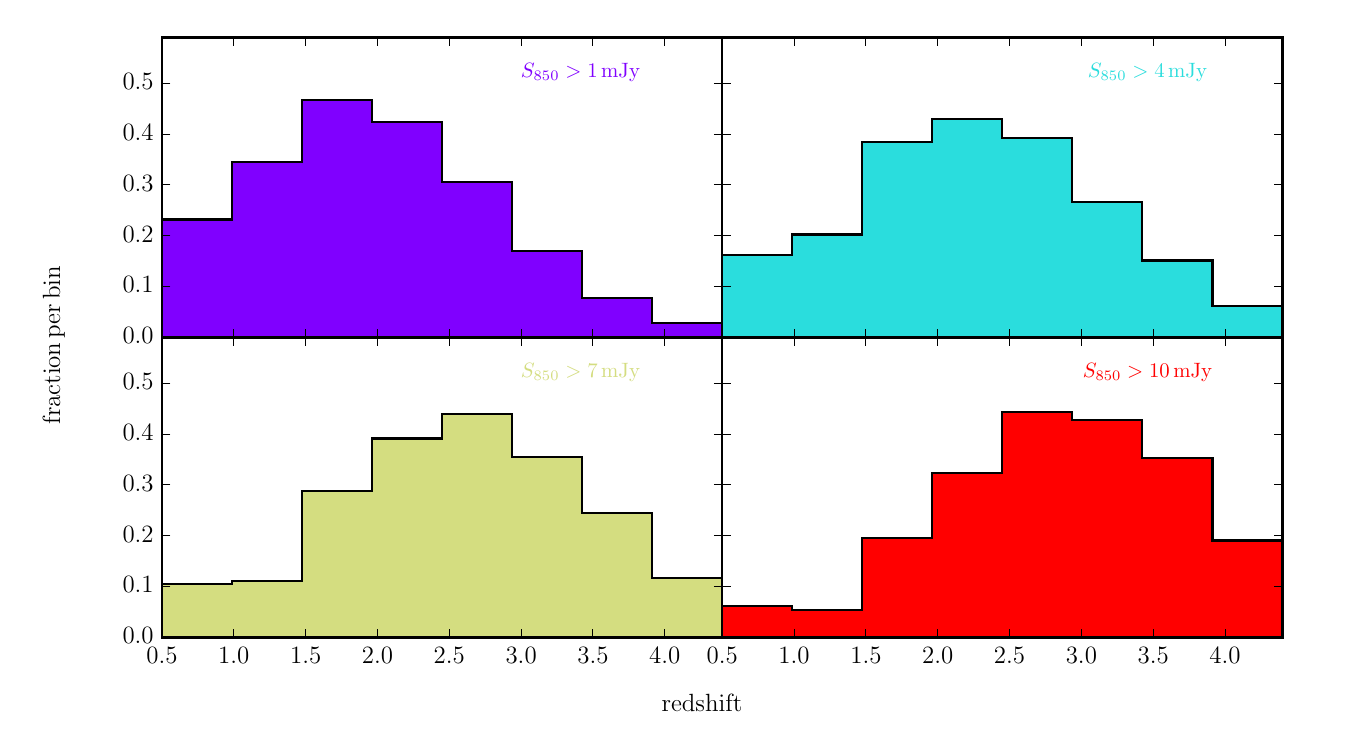}
\end{center}
\caption{The redshift distributions for 850-$\mu$m source samples selected at different limiting flux densities, as predicted from the analytic form of the evolving LF shown in Fig. \ref{fig:lf2}  (Equations \ref{eq:sch} and \ref{eq:params}, with best-fit parameter values listed in Table \ref{tab:params}), derived using the maximum-likelihood method. It can be seen that the brightest sources tend to lie at higher redshifts, with the peak redshift shifting from $z \simeq 1.8$ to $z \simeq 3$ over the flux-density range explored here. This is a direct consequence of the redshift evolution of the IR LF (Figure \ref{fig:lf2}), in particular the continued increase of the characteristic luminosity with redshift. The broad range of redshifts found in each panel explains why it has proved difficult to derive a statistically-significant correlation between $S_{850}$ and $z$, but the predicted relative lack of bright sub-mm sources at $z < 1.5$ accords well with the latest observations (e.g. \citealt{Michalowski_2016}). Although bright 850\,${\rm \mu m}$ sources at $z>3.5$ clearly exist, their relatively low number density, and the overall decline of the LF as seen in Fig. \ref{fig:lf2}, means that their discovery does not conflict with the result that $\rho_{\rm SFR}$ declines beyond $z \simeq 2 - 2.5$ (as seen in Fig. \ref{fig:sfrdc}).}
\label{fig:zcomp}
\end{figure*}

\subsection{Evolution of the LF and the redshift distribution of sub-mm sources}

Finally, we discuss how the form of the LF evolution uncovered here naturally explains the apparent `down-sizing' of the sub-mm source population. There is tentative but persistent evidence that the median redshift of sub-mm selected galaxies increases with increasing flux density \citep{Ivison_2002, Pope_2005, Biggs_2011, Vieira_2013, Weiss_2013, Koprowski_2014, Chen_2016, Michalowski_2016}, and hence increasing far-IR luminosity. In Fig. \ref{fig:lf2} we plot our evolving LFs as a function of redshift from $z \simeq 0.5$ to $z \simeq 4.5$. At low redshifts, the increase in both $\Phi^{\star}$ and $L^{\star}$ produces an increase in both the number density of sources, and luminosity density, $\rho_{\rm IR}$, with increasing redshift (cf Fig. \ref{fig:sfrdc}). At high redshifts the decline in $\Phi^{\star}$ progressively overcomes the continued positive evolution of $L^{\star}$ to produce a decline in both these quantities, but it can be seen that the continued positive evolution of $L^{\star}$ means that the most luminous sources persist, or indeed are preferentially found at the highest redshifts explored here. 

To better connect with observables, we have used our evolving LF to calculate the predicted redshift distribution of 850-$\mu$m sources as a function of flux density. The results are shown for four different flux-density thresholds in Fig.\,12. Here it can be seen that the peak in the redshift distribution is expected to naturally increase gradually from $z \simeq 1.8$ for $S_{850} > 1$\,mJy to $z \simeq 3$ for $S_{850} > 10$\,mJy. This is in excellent accord with what has been reported in the literature (e.g. \citealt{Koprowski_2014, Michalowski_2016}) and clarifies why, although dust-enshrouded star-formation is globally less important than UV-visible star-formation activity at $z > 4$, bright sub-mm sources will continue to be discovered out to high redshifts.

\section{Conclusion}
\label{sec:sum}

We have analysed the coverage of the far-IR luminosity--redshift plane provided by (sub-)mm-selected galaxy samples extracted from the SCUBA-2 Cosmology Legacy Survey and the ALMA imaging of the HUDF to make a new measurement of the evolving galaxy far-IR luminosity function (LF) extending out to redshifts $z \simeq 5$. Using both direct ($1/V_{\rm max}$) and maximum-likelihood methods we have determined the form and evolution of the rest-frame 250-$\mu$m galaxy LF. This LF is well described by a Schechter function with a faint-end slope $\alpha \simeq -0.4$ (derived using the ALMA data at $z \simeq 2$) which displays a combination of rising-then-falling density evolution, and positive luminosity evolution.

We converted our 250-$\mu$m results to total IR luminosity using the average sub-mm galaxy SED of \citet{Michalowski_2010}, and have then compared our determination of the evolving IR LF with those derived by \citet{Magnelli_2013} and \citet{Gruppioni_2013} from the {\it Herschel} PEP and HerMES surveys. Our faint-end slope lies approximately midway between the values adopted in these studies, and the LF normalization near the break luminosity $L^{\star}$ is also comparable at most redshifts. However, both of the {\it Herschel}-derived LFs indicate a much larger number of very bright sources at all redshifts than is found in our JCMT/ALMA-based study. To check which result is correct we derive the predicted 850-$\mu$m number counts from the {\it Herschel}-based LFs, and find that, at bright flux densities ($S_{850} > 10$\,mJy) they predict an order-of-magnitude more sources than are observed. We explore whether this discrepancy can be explained/resolved by adoption of different SED templates, and conclude that it cannot (without extreme temperatures, and more importantly without a contrived and much more rapid dependence of $T$ on IR luminosity than has been reported in the literature).

We have utilised our measurement of the evolving IR LF to derive comoving IR luminosity density, and hence obscured star-formation rate density, which we then combine with UV-estimates of unobscured activity (from \citealt{Parsa_2016}), to derive the evolution of total $\rho_{\rm SFR}$. Again we compare with values derived from the {\it Herschel}-based studies, and find reasonable agreement out to $z \simeq 2$, but increasing disagreement at higher redshift. Specifically, consistent with several other recent studies (e.g. \citealt{Bourne_2017, Dunlop_2017, Liu_2017}) we find that $\rho_{\rm SFR}$ declines beyond $z \simeq 2-2.5$ and is dominated by UV-visible star-formation activity beyond $z \simeq 4$. In contrast, \citet{Gruppioni_2013} and \citet{Rowan_2016} report essentially no decline in dust-obscured $\rho_{\rm SFR}$ from $z \simeq 3$ to $z \simeq 6$. Given the severe over-prediction of the 850-$\mu$m counts produced by the {\it Herschel} IR LFs, we conclude that the high values reported from these studies most likely reflect problems in source identification and redshift estimation arising from the large-beam long-wavelength SPIRE data, as well as potential blending issues, and extrapolation of $\rho_{\rm SFR}$ from the most extreme sources.

Finally, we show how the evolution of the IR LF as derived here (with its combination of rising-then-falling characteristic density ($\Phi^{\star}$), and positive evolution of characteristic luminosity density ($L^{\star}$)) with redshift, produces a decline in inferred $\rho_{\rm SFR}$ beyond $z \simeq 2-2.5$ while at the same time predicting that the most luminous sub-mm sources will continue to be found out to very high redshifts ($z \simeq 5-6$). Specifically, our evolving LF, with its combined luminosity+density evolution, predicts that the median redshift of sub-mm sources should increase with increasing flux density, consistent with several reports in the recent literature. Such evolution is consistent with many studies of AGN evolution, suggesting that both dust-enshrouded star formation and AGN activity are strongly linked to the growth of stellar mass in galaxies.

\section*{Acknowledgements}

KEKC and MPK acknowledge the support of the UK Science and Technology
Facilities Council through grant number ST/M001008/1.
MPK acknowledges the support of the Carnegie Trust Research Incentive Grant (PI: M.~Micha{\l}owski).
JEG is supported by the
Royal Society. JSD,  MJM and RJM acknowledge the support of the UK Science and Technology Facilities Council through grant number ST/M001229/1.
MJM~acknowledges the support of the National Science Centre, Poland
through the POLONEZ grant 2015/19/P/ST9/04010. This project has
received funding from the European Union's Horizon 2020 research and
innovation programme under the Marie Sk{\l}odowska-Curie grant
agreement No.~665778.

The James Clerk Maxwell Telescope is now operated by the
East Asian Observatory on behalf of
The National Astronomical Observatory of Japan, Academia Sinica
Institute of Astronomy and Astrophysics, the Korea Astronomy
and Space Science Institute, the National Astronomical Observa-
tories of China and the Chinese Academy of Sciences (grant no.
XDB09000000), with additional funding support from the Science
and Technology Facilities Council of the United Kingdom and par-
ticipating universities in the United Kingdom and Canada. The data
utilised in this paper were taken as part of Program ID MJLSC02.

This  paper  makes  use  of  the  following  ALMA  data:
ADS/JAO.ALMA\#2012.1.00173.S. ALMA is a partnership of ESO
(representing its member states), NSF (USA) and NINS (Japan), to-
gether with NRC (Canada), NSC and ASIAA (Taiwan), and KASI
(Republic of Korea), in cooperation with the Republic of Chile.
The Joint ALMA Observatory is operated by ESO, AUI/NRAO and
NAOJ.




\bsp	
\label{lastpage}
\end{document}